\documentclass[
 reprint,
 amsmath,
 amssymb,
 aps,
]{revtex4-2}

\usepackage{graphicx}
\usepackage{dcolumn}
\usepackage{bm}

\usepackage{qcircuit}
\usepackage{braket}

\begin{document}

\preprint{APS/123-QED}

\title{
Quantum Algorithm for Radiative Transfer Equation}

\author{Asuka Igarashi$^{1}$}
\email{asuka.igarashi@aist.go.jp, igarashi.asuka.mn@alumni.tsukuba.ac.jp}

\author{Tadashi Kadowaki$^{1,2}$}
\email{tadashi.kadowaki@aist.go.jp}

\author{Shiro Kawabata$^{1,3}$}
\email{Corresponding author, s-kawabata@aist.go.jp}

\affiliation{
$^1$Global Research and Development Center for Business by Quantum-AI technology (G-QuAT), National Institute of Advanced Industrial Science and Technology (AIST), 1-1-1, Umezono, Tsukuba-shi, Ibaraki 305-8568, Japan \\
$^2$DENSO CORPORATION, 1-8-15, Kounan, Minato-ku, Tokyo 108-0075, Japan\\
$^3$NEC-AIST Quantum Technology Cooperative Research Laboratory, National Institute of Advanced Industrial Science and Technology (AIST), 1-1-1, Umezono, Tsukuba-shi, Ibaraki 305-8568, Japan}


\date{\today}

\begin{abstract}
The radiation transfer equation is widely used for simulating such as heat transfer in engineering, diffuse optical tomography in healthcare, and radiation hydrodynamics in astrophysics.
By combining the lattice Boltzmann method, we propose a quantum algorithm for radiative transfer.
This algorithm encompasses all the essential physical processes of radiative transfer: absorption, scattering, and emission.
Although a sufficient number of measurements are required to precisely estimate the quantum state, and the initial encoding of the quantum state remains a challenging problem, our quantum algorithm exponentially accelerates radiative transfer calculations compared to classical algorithms. 
In order to verify the quantum algorithm, we perform quantum circuit simulation using IBM Qiskit Aer and find good agreement between our numerical result and the exact solution. 
The algorithm opens new application of fault-tolerant quantum computers for plasma engineering, telecommunications, nuclear fusion technology, healthcare and astrophysics. 
\end{abstract}
\maketitle


\section{Introduction}

Many studies have utilized the Lattice Boltzmann Method (LBM) as one of the numerical methods in fluid dynamics \cite{McNamara1988, Benzi1992, Higuera1989,Cali1992}.
Notably, the advent of the single-relaxation-time model has significantly developed the field of LBM \cite{Qian1992}. 
Although LBM, which discretizes space, resembles the finite difference method, it does not directly compute macroscopic quantities such as density and velocity.
Instead, LBM calculates particle distribution functions, from which we can derive macroscopic quantities. 
Therefore, LBM is often referred to as a mesoscopic method. 
In addition, many studies developed LBM for various applications such as compressible hydrodynamics \cite{Alexander1993, Yan1999, Kataoka2004a, Kataoka2004b}, thermal hydrodynamic analysis \cite{He1998, Peng2003}, and multiphase flow \cite{Gunstensen1991, Shan1993, Swift1996}. 

Some recent studies have proposed quantum algorithms employing the LBM on fault-tolerant quantum computers (FTQCs) for fluid dynamics \cite{Budinski2021, Itani2023, Sanavio2023, Mezzacapo2015}. 
In the quantum algorithm for LBM, the nonlinearity of the collision term poses a challenge. 
Budinski assumed a low fluid velocity, such that the equilibrium distribution function of the collision term becomes linear, and constructed a quantum algorithm \cite{Budinski2021}.
This algorithm involves two distinctive steps.
The first collision step computes the collision term of the LBM and updates the distribution function.
The circuit depth of this step is independent of the number of lattice points $M$. 
In the second propagation step, the updated distribution function is transferred to adjacent lattice points.
The circuit depth of this step is on the logarithmic order of $M$, indicating the exponential speedup.
A challenge remains for the considerable circuit depth required for the initial encoding \cite{Shende2006}. 
Also, many similar studies proposed various other quantum algorithms for fluid dynamics \cite{Gaitan2020, Li2023, Succi2023b, Steijl2019, Succi2023a} and partial differential equations \cite{Harrow2009, Berry2014, Childs2017, Childs2021, Balducci2022}.

In classical computing, the LBM began to be utilized for radiative transfer in the 2010s \cite{Asinari2010, Ma2011, Bindra2012, Mcculloch2016, Gairola2017, Zhang2013}. 
Radiative transfer is the physical phenomenon of energy transfer in the form of electromagnetic (photon) radiation, and becomes necessary to analyze radiative heat transfer and radiation hydrodynamics. 
The radiative transfer equation takes the form of the Boltzmann equation, allowing for a direct application of the LBM. 
In contrast to conventional fluid particles, electromagnetic waves (photons) do not collide with each other. 
Therefore, the Boltzmann equation for radiative transfer involves only linear terms for the interactions between photons and medium instead of a nonlinear collision term in fluid dynamics. 
The three principal interactions are absorption, scattering, and emission. 
These processes depend on the properties of the medium through which the photons travel and also on the frequency. 
The linearity of radiative transfer avoids the nonlinearity problem for quantum algorithms, such as in fluid dynamics \cite{Steijl2020, Leyton2008, Childs2020}.

Furthermore, radiative transfer has many applications in various other fields. 
Many studies applied radiative transfer in medical imaging, specifically in diffuse optical tomography. 
In diffuse optical tomography, one can determine the condition of bodily tissues from images of near-infrared light that has permeated through the tissues \cite{Gonzalez-Rodriguez2009, Klose1999, Klose2010, Hoshi2016}. 
This technique represents an inverse problem in radiative transfer calculations. 
In the field of astrophysical hydrodynamics, many studies applied radiative transfer analysis to convection of stellar atmospheres \cite{Asplund2000, Hayek2010, Stein1998} and supernova explosions \cite{Blinnikov2000, Piro2010, Hoflich2009, Noebauer2012}, along with the interactions between explosions and circumstellar materials \cite{Kasen2010, Fryer2010}. 
In addition, the radiative transfer equation is used for analyzing telecommunication network \cite{Yahia2021}, soler cell \cite{Gonome2014}, and extreme ultraviolet (EUV) lithography \cite{Su2017}. 
Acceleration of radiative transfer calculations is expected in these industrial fields. 

In this paper, a quantum algorithm for radiative transfer is proposed.
Previous studies show the exponential speedup of quantum algorithms for fluid dynamics based on the LBM \cite{Budinski2021, Li2023, Itani2023, Sanavio2023}.
We demonstrate that such an exponential speedup is also possible for radiative transfer.
Therefore, we can expect fast calculation of radiative transfer phenomena using FTQCs.

The paper is structured as follows.
In Section 2, we transform the radiative transfer equation into the LBM form and modify it to a form suitable for a quantum algorithm.
Section 3 introduces a quantum algorithm for radiative transfer derived from this form.
In this section, we show quantum circuits.
In Section 4, we compute a simple one-dimensional problem using the quantum algorithm and compare it with the classical algorithm and analytical solutions.
Sections 5 and 6 encompass discussions and conclusions, respectively.

\section{Formulation}
\begin{figure}[htbp]
    \centering
    \includegraphics[scale=0.11]{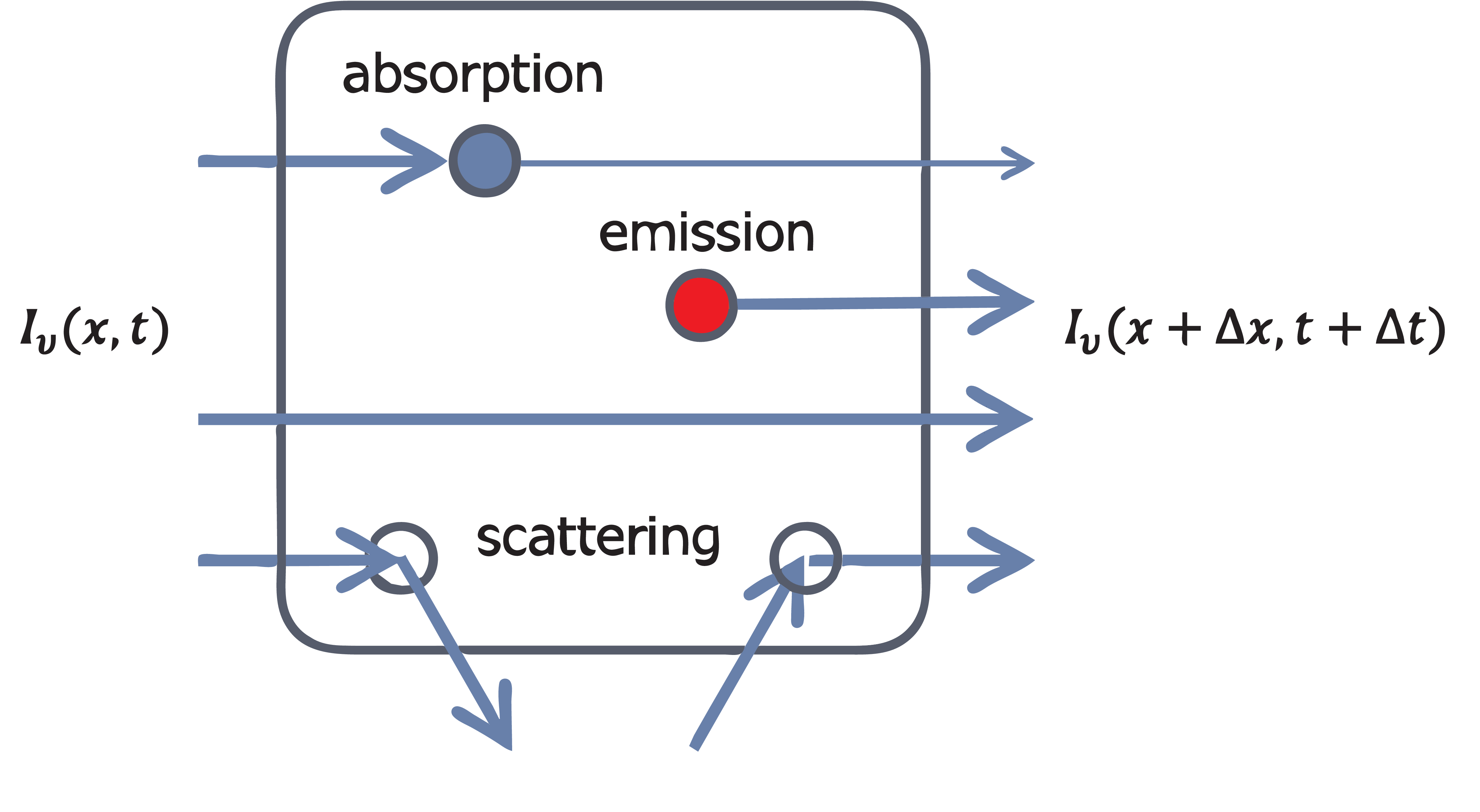}
    \caption{
    Essential physical processes for radiative transfer.}
    \label{figure_radiative_transfer}
\end{figure}

Radiative transfer (also known as radiative transport) is the physical phenomenon in which electromagnetic waves (photons) transmit energy through a medium \cite{Chandrasekhar2011}. 
Radiation propagation includes processes such as absorption, emission, and scattering (see Fig. \ref{figure_radiative_transfer}). 
The radiative transfer equation mathematically describes these processes. 
Analytical solutions to this equation exist only for simple cases, and more realistic cases require numerical methods. 
The radiative transfer equation is given by an integro-differential equation: 
\begin{eqnarray}
&&\frac{\partial I_{\nu}(\vec{x}, \vec{\Omega}, t)}{c \partial t} 
+ \vec{\Omega} \cdot \nabla I_{\nu}(\vec{x}, \vec{\Omega}, t) \nonumber\\
&& \quad 
= \kappa_{a,\nu}(\vec{x}, \vec{\Omega}, t) \left\{ I_{b,\nu}(\vec{x}, \vec{\Omega}, t) - I_{\nu}(\vec{x}, \vec{\Omega}, t) \right\} \nonumber\\
&& \quad \quad 
+ \sigma_{s,\nu}(\vec{x}, \vec{\Omega}, t) \nonumber\\
&& \quad \quad \quad 
\times \left\{ \int_{4\pi} I_{\nu}(\vec{x}, \vec{\Omega'}, t) \Phi(\vec{\Omega},\vec{\Omega}') \partial\vec{\Omega}' - I_{\nu}(\vec{x}, \vec{\Omega}, t) \right\} \nonumber\\
&& \quad \quad  
+ S_{e,\nu}(\vec{x}, \vec{\Omega}, t),
\label{general_radiative_transfer_equation}
\end{eqnarray}
where $c$, $\vec{x}$, $t$, $\vec{\Omega}$, and $I_{\nu}$ represent the speed of light,  position, time, angular direction, and radiative intensity at frequency $\nu$, respectively. 
On the right-hand side, $\kappa_{a,\nu}$, $\sigma_{s,\nu}$, $S_{e,\nu}$, $I_{b,\nu}$ and $\Phi$ represent the absorption coefficient, the scattering coefficient, the emission source, the black-body radiative intensity, and the scattering phase function.
The left-hand side of Eq. (\ref{general_radiative_transfer_equation}) describes time evolution of the radiative intensity $I_{\nu}$, similar to the Boltzmann equation for fluid dynamics.
The right-hand side represents changes by various physical processes. 
The first term describes absorption, the second term for scattering, and the third term for emission. 
In contrast to the Boltzmann equation for fluid dynamics, which contains the nonlinear collision term, the radiative transfer equation has linear terms only. 

We assume the isotropic scattering and modify the general form (\ref{general_radiative_transfer_equation}) to the tractable form:
\begin{eqnarray}
&&\frac{\partial I_{\nu}(\vec{x}, \vec{\Omega}, t)}{c \partial t} 
+ \vec{\Omega} \cdot \nabla I_{\nu}(\vec{x}, \vec{\Omega}, t) \nonumber\\
&& \quad 
= -\kappa_{\nu}(\vec{x}, \vec{\Omega}, t) I_{\nu}(\vec{x}, \vec{\Omega}, t) \nonumber\\
&& \quad \quad 
+ \sigma_{\nu}(\vec{x}, \vec{\Omega}, t) \int_{4\pi} I_{\nu}(\vec{x}, \vec{\Omega}', t) \partial \vec{\Omega}' 
+ S_{\nu}(\vec{x}, \vec{\Omega}, t). 
\label{simple_radiative_transfer_equation}
\end{eqnarray}
The coefficient $\kappa_{\nu} (\equiv \kappa_{a,\nu} + \sigma_{s,\nu})$ denotes a decrease of $I_{\nu}$ due to absorption and scattering, while $\sigma_{\nu} (\equiv \sigma_{s,\nu} / 4\pi)$ denotes an increase of $I_{\nu}$ due to scattering.
The source term $S_{\nu} (\equiv \kappa_{a,\nu} I_{b,\nu} + S_{e,\nu})$ represents an increase due to absorption and emission. 

\begin{figure}[htbp]
    \centering
    \includegraphics[scale=0.13]{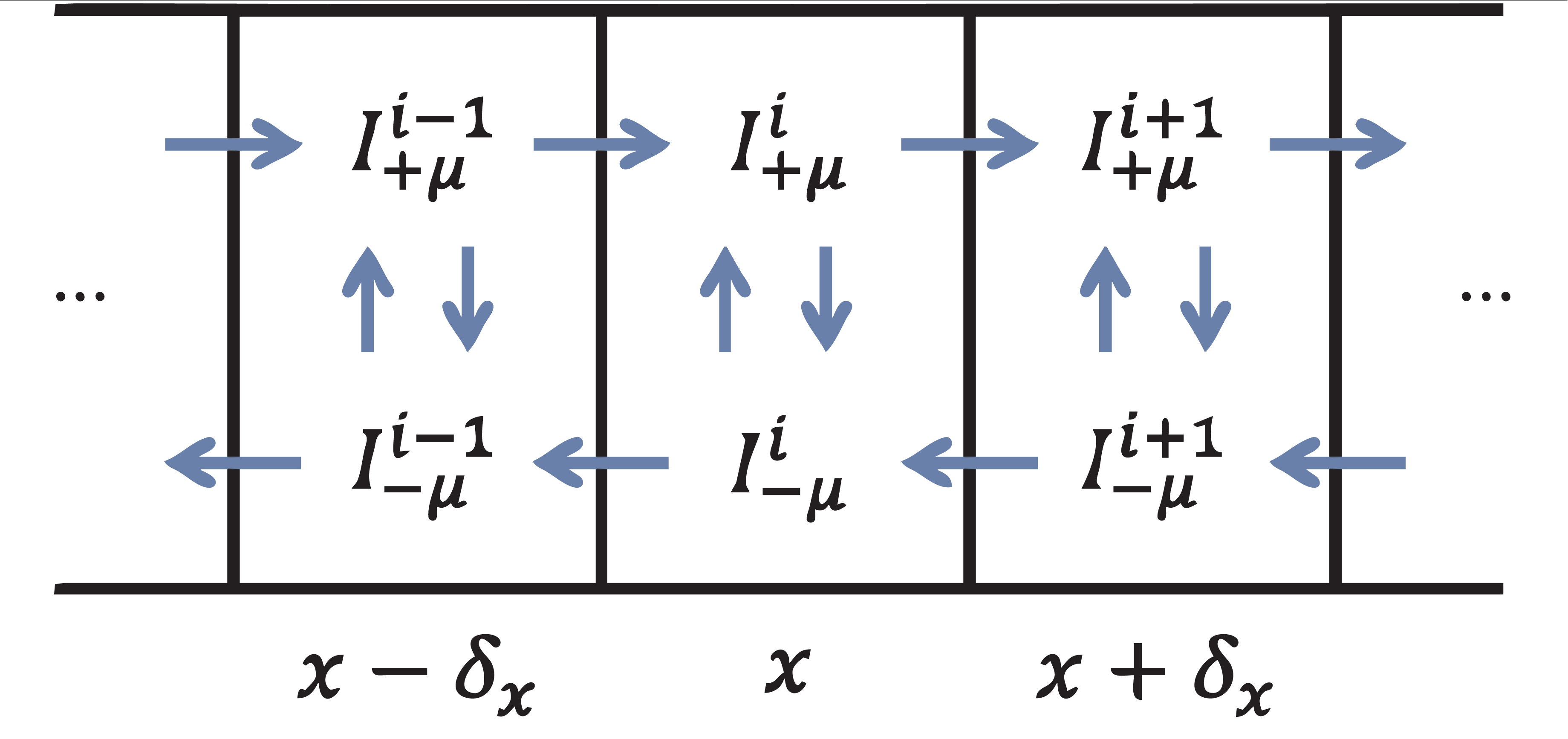}
    \caption{
    One-dimensional model for radiative transfer.}
    \label{figure_one_dimensional_model}
\end{figure}

In this paper, we propose the quantum algorithm for the LBM of radiative transfer. 
We show the quantum circuit for a simple case, which is a one-dimensional radiative transfer problem (see Fig. \ref{figure_one_dimensional_model}). 
This model is similar to the D1Q2 model (one-dimensional flow with two discrete velocities) in the LBM-based fluid dynamics \cite{Succi2001}. 
Equation (\ref{simple_radiative_transfer_equation}) can be transformed into the one-dimensional LBM form:
\begin{eqnarray}
&& I_{\pm \mu} (x \pm \delta_x, t + \delta_t) - I_{\pm \mu} (x, t) \nonumber\\
&& \quad = -\kappa\delta_t I_{\pm \mu} (x, t)
+ \frac{\sigma}{2}\delta_t \left\{ I_{+\mu}(x, t) + I_{-\mu}(x, t) \right\} \nonumber\\
&& \qquad \quad + \frac{1}{2} \delta_tS_{\pm \mu} (x),
\label{lattice_boltzmann_equation}
\end{eqnarray}
where $\pm\mu$ indicates the two angular directions of radiative transfer, $\delta_t$ and $\delta_x (\equiv c\mu\delta_t)$ represent the time step and the grid width between lattice points. 
For simplicity, we have assumed that each coefficient ($\kappa$, $\sigma$) is constant and has a fixed frequency $\nu$.
In equation (\ref{lattice_boltzmann_equation}), $1/2$ is the weight factor due to the presence of two directions. 

In order to develop the quantum algorithm for the lattice Boltzmann equation (\ref{lattice_boltzmann_equation}), we decompose this equation as:
\begin{eqnarray}
I_{\pm\mu} (x \pm \delta_x, t + \delta_t) 
= I^{ASE}_{\pm\mu} (x,t) ,
\label{equation_propagation_step}
\end{eqnarray}
\begin{eqnarray}
I^{ASE}_{\pm\mu} (x,t) 
\equiv I^{AS}_{\pm\mu} (x,t) + \frac{1}{2} S_{\pm\mu} (x)\delta_t ,
\label{equation_absorption_and_emission_step}
\end{eqnarray}
where
\begin{eqnarray}
\left\{
\renewcommand{\arraystretch}{1.5}
\begin{array}{lll}
I^{AS}_{+\mu} (x,t) 
&\equiv& \left( 1 - \kappa\delta_t + \frac{\sigma}{2}\delta_t \right) I_{+\mu} (x,t)  \\
&\qquad& \qquad + \frac{\sigma}{2}\delta_t I_{-\mu} (x,t), \\
I^{AS}_{-\mu} (x,t) 
&\equiv& \frac{\sigma}{2}\delta_t I_{+\mu} (x,t) \\
&\qquad& \qquad + \left( 1 - \kappa\delta_t + \frac{\sigma}{2}\delta_t \right) I_{-\mu} (x,t). 
\end{array}
\right. 
\label{equation_absorption_and_scattering_step}
\end{eqnarray}
The above equations indicate that the quantum algorithm involves three steps: an absorption-and-scattering step ($I_{\pm\mu} \rightarrow I^{AS}_{\pm\mu}$) as Eq. (\ref{equation_absorption_and_scattering_step}), an absorption-and-emission step ($I^{AS}_{\pm\mu} \rightarrow I^{ASE}_{\pm\mu}$) as Eq. (\ref{equation_absorption_and_emission_step}), and a propagation step ($I^{ASE}_{\pm\mu} \rightarrow I_{\pm\mu}$) as Eq. (\ref{equation_propagation_step}). 

\section{Quantum Algorithm}

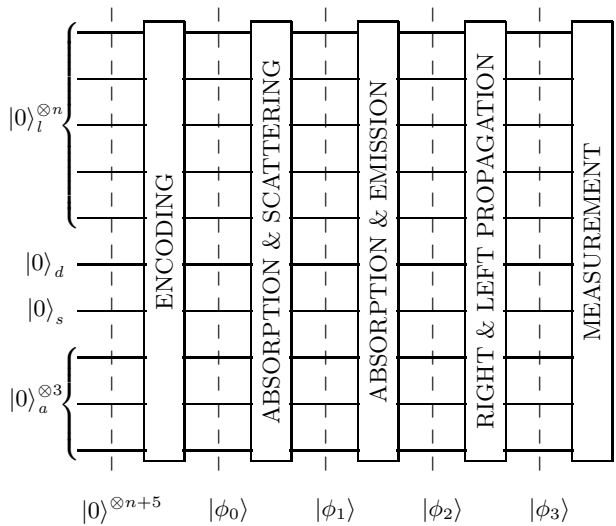
\begin{figure}[htbp]
    \centering
    \begin{tabular}{lr} 
        \hspace{8mm} &
        \Qcircuit @C=0.9em @R=1em {
             & \qw \ar@{--}[]+<0.5em,1em>;[d]+<0.5em,-16em> & \qw &
             \multigate{9}{\rotatebox{90}{\small ENCODING}}                  & 
             \qw \ar@{--}[]+<0.5em,1em>;[d]+<0.5em,-16em> & \qw &
             \multigate{9}{\rotatebox{90}{\small ABSORPTION \& SCATTERING}}  & 
             \qw \ar@{--}[]+<0.5em,1em>;[d]+<0.5em,-16em> & \qw &
             \multigate{9}{\rotatebox{90}{\small ABSORPTION \& EMISSION}}    & 
             \qw \ar@{--}[]+<0.5em,1em>;[d]+<0.5em,-16em> & \qw &
             \multigate{9}{\rotatebox{90}{\small RIGHT \& LEFT PROPAGATION}} & 
             \qw \ar@{--}[]+<0.5em,1em>;[d]+<0.5em,-16em> & \qw &
             \multigate{9}{\rotatebox{90}{\small MEASUREMENT}} 
             \inputgroupv{1}{5}{0.7em}{3.5em}{\ket{{\rm 0}}_{\it l}^{\it \otimes n}}\\
             & \qw & \qw & \ghost{1} & \qw & \qw & \ghost{1} & \qw & \qw & \ghost{1} & \qw & \qw & \ghost{1} & \qw & \qw & \ghost{1}  \\
             & \qw & \qw & \ghost{1} & \qw & \qw & \ghost{1} & \qw & \qw & \ghost{1} & \qw & \qw & \ghost{1} & \qw & \qw & \ghost{1}  \\
             & \qw & \qw & \ghost{1} & \qw & \qw & \ghost{1} & \qw & \qw & \ghost{1} & \qw & \qw & \ghost{1} & \qw & \qw & \ghost{1}  \\
             & \qw & \qw & \ghost{1} & \qw & \qw & \ghost{1} & \qw & \qw & \ghost{1} & \qw & \qw & \ghost{1} & \qw & \qw & \ghost{1}  \\
             \lstick{\ket{{\rm 0}}_{\it d}} 
             & \qw & \qw & \ghost{1} & \qw & \qw & \ghost{1} & \qw & \qw & \ghost{1} & \qw & \qw & \ghost{1} & \qw & \qw & \ghost{1}  \\
             \lstick{\ket{{\rm 0}}_{\it s}} 
             & \qw & \qw & \ghost{1} & \qw & \qw & \ghost{1} & \qw & \qw & \ghost{1} & \qw & \qw & \ghost{1} & \qw & \qw & \ghost{1}  \\
             & \qw & \qw & \ghost{1} & \qw & \qw & \ghost{1} & \qw & \qw & \ghost{1} & \qw & \qw & \ghost{1} & \qw & \qw & \ghost{1}  
             \inputgroupv{8}{10}{0.7em}{1.7em}{\ket{{\rm 0}}_{\it a}^{\otimes 3}} \\
             & \qw & \qw & \ghost{1} & \qw & \qw & \ghost{1} & \qw & \qw & \ghost{1} & \qw & \qw & \ghost{1} & \qw & \qw & \ghost{1}  \\
             & \qw & \qw & \ghost{1} & \qw & \qw & \ghost{1} & \qw & \qw & \ghost{1} & \qw & \qw & \ghost{1} & \qw & \qw & \ghost{1}  \\
             & & & & & & & & & & & & & & & & \\
             & & \ket{0}^{\otimes n+5} & & & \ket{\phi_0} & & & \ket{\phi_1} & & & \ket{\phi_2} & & & \ket{\phi_3} & & 
        }
    \end{tabular}
    \caption{Quantum circuit for all steps.
    The brackets $\ket{0}_{\it l}^{\it \otimes n}$, $\ket{0}_{\it d}$, $\ket{0}_{\it s}$, and $\ket{0}_{\it a}^{\rm \otimes 3}$ represent lattice qubits, a direction qubit, a switching qubit, and ancilla qubits, respectively.}
    \label{circuit_total_step}
\end{figure}

The quantum algorithm consists of the following four steps : the encoding step, the absorption-and-scattering step, the absorption-and-emission step, and the propagation step. 
We summarize all steps of the quantum algorithm in Fig. \ref{circuit_total_step}. 
First, we encode the classical state $I_{\pm\mu}(x,t)$ of Eq. (\ref{lattice_boltzmann_equation}) into the quantum state. 
We define the classical state vector $\vec{\phi}$ as 
\begin{eqnarray}
    \vec{\phi} 
    & \equiv &  
    \biggl( 
    I_{+\mu}^0, \cdots, I_{+\mu}^{M-1},
    I_{-\mu}^0, \cdots, I_{-\mu}^{M-1}, 
    \nonumber\\
    & & \quad \quad \quad 
    \frac{1}{2} \delta_t S_{+\mu}^0, \cdots, \frac{1}{2} \delta_t S_{+\mu}^{M-1}, \nonumber\\
    & & \quad \quad \quad \quad \quad 
    \left.
    \frac{1}{2} \delta_t S_{+\mu}^0, \cdots, \frac{1}{2} \delta_t S_{+\mu}^{M-1} 
    \right),
\end{eqnarray}
where $M$ is the number of lattices and the superscript ($0\sim M-1$) denotes the lattice site. 
The encoded quantum state is 
\begin{eqnarray}
    \ket{\phi_0} 
    & = &
    \ket{000}_{\it a} 
    \nonumber\\
    &&
    \times \frac{1}{\|\phi\|} 
    \sum_{i=0}^{M-1} 
    \biggl(
      I_{+\mu}^i \ket{0}_{\it s} \ket{0}_{\it d} \ket{i}_{\it l}
    + I_{-\mu}^i \ket{0}_{\it s} \ket{1}_{\it d} \ket{i}_{\it l}
    \nonumber\\
    && \quad \quad \quad \quad \quad \quad 
    + \frac{1}{2} \delta_t S_{+\mu}^i \ket{1}_{\it s} \ket{0}_{\it d} \ket{i}_{\it l}
    \nonumber\\
    && \quad \quad \quad \quad \quad \quad \quad \quad 
    \left.
    + \frac{1}{2} \delta_t S_{-\mu}^i \ket{1}_{\it s} \ket{1}_{\it d} \ket{i}_{\it l}
    \right),
    \label{quantum_state_0}
\end{eqnarray}
where $i$ denotes the binary representation of the lattice site ($=0\sim M-1$). 
We scale amplitude by the norm $\|\phi\|$ for normalization. 
The three qubits labeled as $a$ represent ancillary qubits. 
In this quantum algorithm, we have to measure the state after the each time step (Fig. \ref{circuit_total_step}) and the ancillary qubits are indexes to distinguish the desired amplitudes. 
The $s$-labeled qubit indicates $I$ and $S$, where 0 corresponds to $I$ and 1 corresponds to $S$. 
The $d$-labeled qubit distinguishes left and right, where 0 represents the $+\mu$ direction, and 1 represents the $-\mu$ direction. 
The $n$ qubits labeled as $l$ denote lattice points ($M=2^n$,$i\in \{0,1\}^{\otimes n}$). 
As a result, we need $n+5$ qubits for the quantum algorithm. 

From equations (\ref{equation_absorption_and_scattering_step}), the calculations for the absorption-and-scattering step ($I_{\pm\mu}\rightarrow I_{\pm\mu}^{\rm AS}$) are as follows:
\begin{eqnarray}
    &&
    \begin{pmatrix}
        I_{+\mu}^{AS,0}, \cdots, I_{+\mu}^{AS,M-1}, I_{-\mu}^{AS,0}, \cdots, I_{-\mu}^{AS,M-1}
    \end{pmatrix}^T
    \nonumber\\
    && \quad \quad \quad \quad = A
    \begin{pmatrix}
        I_{+\mu}^{0},    \cdots, I_{+\mu}^{M-1},    I_{-\mu}^{0},    \cdots, I_{-\mu}^{M-1}
    \end{pmatrix}^T.
\end{eqnarray}
In this equation, 
\begin{eqnarray}
    A\equiv
    \renewcommand{\arraystretch}{1.5}
    \begin{pmatrix}
        1 - \kappa \delta_t + \frac{1}{2} \sigma \delta_t & \frac{1}{2} \sigma \delta_t \\
        \frac{1}{2} \sigma \delta_t & 1 - \kappa \delta_t+\frac{1}{2} \sigma\delta_t
    \end{pmatrix}
    \otimes I_M,
\end{eqnarray}
where $A$ is the $2M\times 2M$ non-unitary matrix and $I_M$ is $M\times M$ identity matrix.
We utilize the linear combination of unitaries (LCU) method employed in \cite{Budinski2021}. 
This approach decomposes the $2M\times 2M$ non-unitary matrix $A$ into two $2M\times 2M$ unitary matrices, denoted as $C_{1}$ and $C_{2}$, i.e., $A= (C_1 + C_2)/2$, where
\begin{eqnarray}
    C_{1,2} 
    &\equiv& A \pm \frac{1}{2} i \sqrt{I-A^2} 
    \nonumber\\
    &=&
    \renewcommand{\arraystretch}{1.5}
    \begin{pmatrix}
        a_0 \pm \frac{1}{2} i b_0 & a_1 \pm \frac{1}{2} i b_1 \\
        a_1 \pm \frac{1}{2} i b_1 & a_0 \pm \frac{1}{2} i b_0
    \end{pmatrix} 
    \otimes I_M.
\end{eqnarray}
In this equation, 
\begin{eqnarray}
    \left\{
    \renewcommand{\arraystretch}{1.5}
    \begin{array}{lll}
        a_0     & \equiv & 1 - \kappa \delta_t + \frac{1}{2} \sigma \delta_t,\\
        a_1     & \equiv & \frac{1}{2} \sigma \delta_t, \\
        b_{0,1} & \equiv & \sqrt{1 - (1 - \kappa \delta_t + \sigma \delta_t)^2} \\
        & & \quad \quad \quad \quad \quad \quad \quad \quad 
        \pm \sqrt{1 - (1 - \kappa \delta_t)^2 }, 
    \end{array}
    \right.
\end{eqnarray}
where $a_{0,1}$ and $b_{0,1}$ are dimensionless constants. 
The two matrices $C_1$ and $C_2$ can be decomposed using $X$-gate, $P$-gate (phase gate), and $R_X$-gate as follows:
\begin{eqnarray}
    C_{1,2} 
    &=& e^{i\alpha_{1,2}} R_X(\beta_{1,2}) \otimes I_M\nonumber\\
    &=& X P(\alpha_{1,2}) X P(\alpha_{1,2}) R_{\rm X}(\beta_{1,2}) \otimes I_M, 
\end{eqnarray}
where $\alpha_{1,2}$ and $\beta_{1,2}$ are real constants.
Here, $X$, $P$ and $R_X$ gates are
\begin{eqnarray}
    &&
    X  = 
    \begin{pmatrix}
        0 & 1 \\
        1 & 0
    \end{pmatrix},
    \\
    &&
    P (\alpha)  = 
    \begin{pmatrix}
        1 & 0 \\
        0 & e^{i\alpha}
    \end{pmatrix}, 
    \\
    &&
    R_X (\beta)  = 
    \begin{pmatrix}
          \cos \left(\beta /2\right) & -i\sin \left(\beta /2\right) \\
        -i\sin \left(\beta /2\right) &   \cos \left(\beta /2\right)
    \end{pmatrix}.
\end{eqnarray}
We can determine $\alpha_{1,2}$ and $\beta_{1,2}$ from $a_{0,1}$ and $b_{0,1}$ by solving: 
\begin{eqnarray}
    \left\{
    \renewcommand{\arraystretch}{1.5}
    \begin{array}{rcl}
          e^{i\alpha_1} \cos \frac{\beta_1}{2} & = & a_0 + \frac{1}{2} i b_0, \\
        -ie^{i\alpha_1} \sin \frac{\beta_1}{2} & = & a_1 + \frac{1}{2} i b_1, \\
          e^{i\alpha_2} \cos \frac{\beta_2}{2} & = & a_0 - \frac{1}{2} i b_0, \\
        -ie^{i\alpha_2} \sin \frac{\beta_2}{2} & = & a_1 - \frac{1}{2} i b_1.
    \end{array}
    \right.
\end{eqnarray}
It should be noted that $b_{0,1}$ can be complex. 

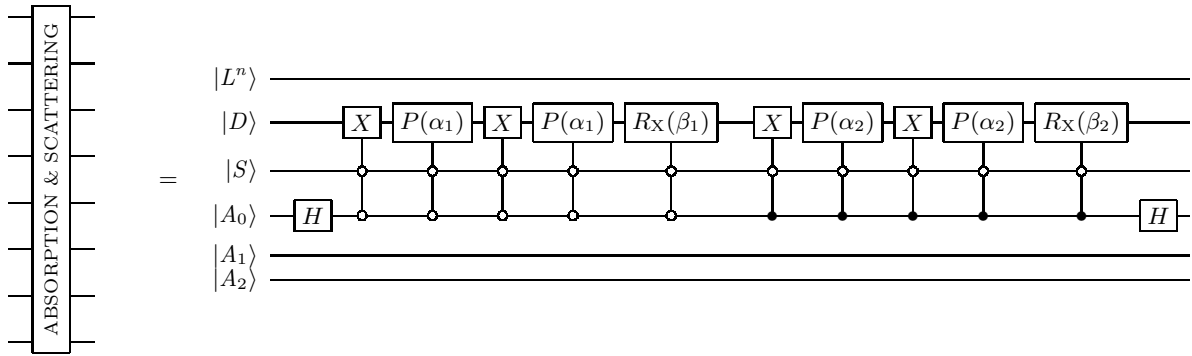
\begin{figure*}[htbp]
    \begin{tabular}{ccc}
        \begin{minipage}[h]{0.45\linewidth}
            \centering
            \Qcircuit @C=1em @R=1em {
                 & \multigate{7}{\rotatebox{90}{\scriptsize ABSORPTION \& SCATTERING}} & \qw  \\
                 & \ghost{1} & \qw \\
                 & \ghost{1} & \qw \\
                 & \ghost{1} & \qw \\
                 & \ghost{1} & \qw \\
                 & \ghost{1} & \qw \\
                 & \ghost{1} & \qw \\
                 & \ghost{1} & \qw
            }
        \end{minipage} &
        \: \: \: \: = \: \: \: \: \: \: &
        \begin{minipage}[h]{0.45\linewidth}
            \centering
            \Qcircuit @C=0.5em @R=1em {
                \lstick{\ket{{\it L^n}}} & \qw & \qw      & \qw        & \qw                & \qw        & \qw                & \qw                 & \qw & \qw & \qw        & \qw                & \qw        & \qw                & \qw                 & \qw      & \qw \\
                \lstick{\ket{{\it D}}}   & \qw & \qw      & \gate{X}   & \gate{P(\alpha_1)} & \gate{X}   & \gate{P(\alpha_1)} & \gate{R_{\rm X}(\beta_1)} & \qw & \qw & \gate{X}   & \gate{P(\alpha_2)} & \gate{X}   & \gate{P(\alpha_2)} & \gate{R_{\rm X}(\beta_2)} & \qw      & \qw \\
                \lstick{\ket{{\it S}}}   & \qw & \qw      & \ctrlo{-1} & \ctrlo{-1}         & \ctrlo{-1} & \ctrlo{-1}         & \ctrlo{-1}          & \qw & \qw & \ctrlo{-1} & \ctrlo{-1}         & \ctrlo{-1} & \ctrlo{-1}         & \ctrlo{-1}          & \qw      & \qw \\
                \lstick{\ket{{\it A}_0}} & \qw & \gate{H} & \ctrlo{-1} & \ctrlo{-1}         & \ctrlo{-1} & \ctrlo{-1}         & \ctrlo{-1}          & \qw & \qw & \ctrl{-1}  & \ctrl{-1}          & \ctrl{-1}  & \ctrl{-1}          & \ctrl{-1}           & \gate{H} & \qw \\
                \lstick{\ket{{\it A}_1}} & \qw & \qw      & \qw        & \qw                & \qw        & \qw                & \qw                 & \qw & \qw & \qw        & \qw                & \qw        & \qw                & \qw                 & \qw      & \qw \\
                \lstick{\ket{{\it A}_2}} & \qw & \qw      & \qw        & \qw                & \qw        & \qw                & \qw                 & \qw & \qw & \qw        & \qw                & \qw        & \qw                & \qw                 & \qw      & \qw
            }
        \end{minipage}
    \end{tabular}
    \caption{Quantum circuit for the absorption-and-scattering step.
    The state vectors $\ket{\it L^n}$, $\ket{\it D}$, $\ket{\it S}$, $\ket{{\it A}_0}$, $\ket{{\it A}_1}$, and $\ket{{\it A}_2}$ represent lattice qubits ($n$ qubits), a direction qubit, a switching qubit, and three ancilla qubits, respectively.
    Determination of the parameters $\alpha_{1,2}$ and $\beta_{1,2}$ inside the gates are described in the main text.}
    \label{circuit_absorption_step}
\end{figure*}

Fig. \ref{circuit_absorption_step} shows the quantum circuit for the absorption-and-scattering step. 
Applying this circuit to $\ket{\phi_0}$ results in a quantum state: 
\begin{eqnarray}
    \ket{\phi_1} 
    & = &
    \ket{000}_{\it a} 
    \nonumber\\
    &&
    \times \frac{1}{\|\phi\|}
    \sum_{i=0}^{M-1}
    \biggl( 
      I_{+\mu}^{AS,i} \ket{0}_{\it s}\ket{0}_{\it d} \ket{i}_{\it l}
    + I_{-\mu}^{AS,i} \ket{0}_{\it s}\ket{1}_{\it d} \ket{i}_{\it l}
    \nonumber\\
    && \quad
    \left.
    + \frac{1}{2} \delta_t S_{+\mu}^i \ket{1}_{\it s}\ket{0}_{\it d} \ket{i}_{\it l}
    + \frac{1}{2} \delta_t S_{-\mu}^i \ket{1}_{\it s}\ket{1}_{\it d} \ket{i}_{\it l}
    \right) \nonumber\\
    &&
    + \ket{000^{\perp}}_{\it a,s,d,l} ,
    \label{quantum_state_1}
\end{eqnarray}
where $\ket{000^{\perp}}_{a,s,d,l}$ represents the component orthogonal to $\ket{000}_a$. 
Because we require the amplitude of $\ket{000}_a$, $\ket{000^{\perp}}_{a,s,d,l}$ is not of interest in this calculation. 

\begin{figure}[htbp]
    \begin{tabular}{ccc}
        \begin{minipage}[h]{0.45\linewidth}
            \centering
            \Qcircuit @C=1em @R=1em {
                 & \multigate{6}{\rotatebox{90}{\scriptsize ABSORPTION \& EMISSION}} & \qw  \\
                 & \ghost{1} & \qw \\
                 & \ghost{1} & \qw \\
                 & \ghost{1} & \qw \\
                 & \ghost{1} & \qw \\
                 & \ghost{1} & \qw \\ 
                 & \ghost{1} & \qw 
            }
        \end{minipage} &
        \: \: = \: \: \: \: \: \: &
        \begin{minipage}[h]{0.45\linewidth}
            \centering
            \Qcircuit @C=1em @R=1em {
                \lstick{\ket{{\it L^n}}} & \qw & \qw      & \qw        & \qw       & \qw       & \qw      & \qw \\
                \lstick{\ket{{\it D}}}   & \qw & \qw      & \qw        & \qw       & \qw       & \qw      & \qw \\
                \lstick{\ket{{\it S}}}   & \qw & \qw      & \gate{X}   & \gate{X}  & \gate{Z}  & \qw      & \qw \\
                \lstick{\ket{{\it A}_0}} & \qw & \qw      & \qw        & \qw       & \qw       & \qw      & \qw \\
                \lstick{\ket{{\it A}_1}} & \qw & \gate{H} & \ctrl{-2}  & \ctrl{-2} & \ctrl{-2} & \gate{H} & \qw \\
                \lstick{\ket{{\it A}_2}} & \qw & \gate{H} & \ctrlo{-1} & \ctrl{-1} & \ctrl{-1} & \gate{H} & \qw 
            }
        \end{minipage}
    \end{tabular}
    \caption{Quantum circuit for the absorption-and-emission step.
    The symbols within the brackets are the same as in Figure \ref{circuit_absorption_step}.}
    \label{circuit_emission_step}
\end{figure}
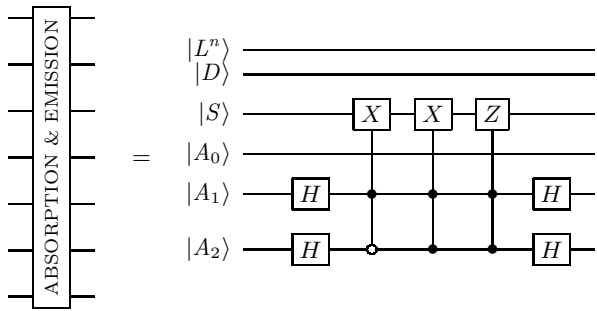

Next, we proceed to compute the absorption-and-emission step. 
This step can be expressed using Eq. (\ref{equation_absorption_and_emission_step}) as follows: 
\begin{eqnarray}
    &&
    \biggl(
    I_{+\mu}^{ASE,0}, \cdots, I_{+\mu}^{ASE,M-1}, I_{-\mu}^{ASE,0}, \cdots, I_{-\mu}^{ASE,M-1}, \nonumber\\
    && \:
    \frac{1}{2} \delta_t S_{+\mu}^{0}, \cdots, \frac{1}{2} \delta_t S_{+\mu}^{M-1}, 
    \left. 
    \frac{1}{2} \delta_t S_{-\mu}^{0}, \cdots, \frac{1}{2} \delta_t S_{-\mu}^{M-1} 
    \right)^T
    \nonumber\\
    && = B
    \biggl(
    I_{+\mu}^{AS,0}, \cdots, I_{+\mu}^{AS,M-1}, I_{-\mu}^{AS,0}, \cdots, I_{-\mu}^{AS,M-1}, 
    \nonumber\\
    && \: \quad \quad \quad \quad 
    \frac{1}{2} \delta_t S_{+\mu}^{0}, \cdots, \frac{1}{2} \delta_t S_{+\mu}^{M-1}, 
    \nonumber\\
    && \: \quad \quad \quad \quad \quad \quad 
    \left. 
    \frac{1}{2} \delta_t S_{-\mu}^{0}, \cdots, \frac{1}{2} \delta_t S_{-\mu}^{M-1} 
    \right)^T. 
\end{eqnarray}
In this equation, 
\begin{eqnarray}
    B 
    &\equiv& 
    \begin{pmatrix}
        I_{2M} & I_{2M} \\
        0 & I_{2M}
    \end{pmatrix}
    \nonumber\\
    &=& 
    \left( I + \frac{1}{2} X + \frac{1}{2} ZX \right) \otimes I_{2M}, 
\end{eqnarray}
where $B$ is the $4M\times 4M$ non-unitary matrix and $I_{2M}$ is $2M\times 2M$ identity matrix. 
We compute $B$ using a combination of $X$-gate and $Z$-gate with two ancillary qubits. 
The $Z$-gate is
\begin{eqnarray}
    Z = 
    \begin{pmatrix}
        1 & 0 \\
        0 & -1
    \end{pmatrix}.
\end{eqnarray}
Fig. \ref{circuit_emission_step} shows the quantum circuit for the absorption-and-emission step. 
By applying this circuit to $\ket{\phi_1}$, the resulting quantum state is given by: 
\begin{eqnarray}
    \ket{\phi_2}
    &=&
    \ket{000}_{\it a} 
    \nonumber\\
    && \: 
    \times \frac{1}{2\|\phi\|}
    \sum_{i=0}^{M-1} 
    \biggl(
      I_{+\mu}^{ASE,i} \ket{0}_{\it s}\ket{0}_{\it d} \ket{i}_{\it l}
    \nonumber\\
    && \: \quad 
    + I_{-\mu}^{ASE,i} \ket{0}_{\it s} \ket{1}_{\it d} \ket{i}_{\it l} 
    + \frac{1}{2} \delta_t S_{+\mu}^i \ket{1}_{\it s} \ket{0}_{\it d} \ket{i}_{\it l} 
    \nonumber\\
    && \: \quad 
    \left.
    + \frac{1}{2} \delta_t S_{-\mu}^i \ket{1}_{\it s} \ket{1}_{\it d} \ket{i}_{\it l}
    \right) 
    + \ket{000^{\perp}}'_{\it a,s,d,l} , 
    \label{quantum_state_2}
\end{eqnarray}
where again $\ket{000^{\perp}}'_{\it a,s,d,l}$ is irrelevant for our purpose. 

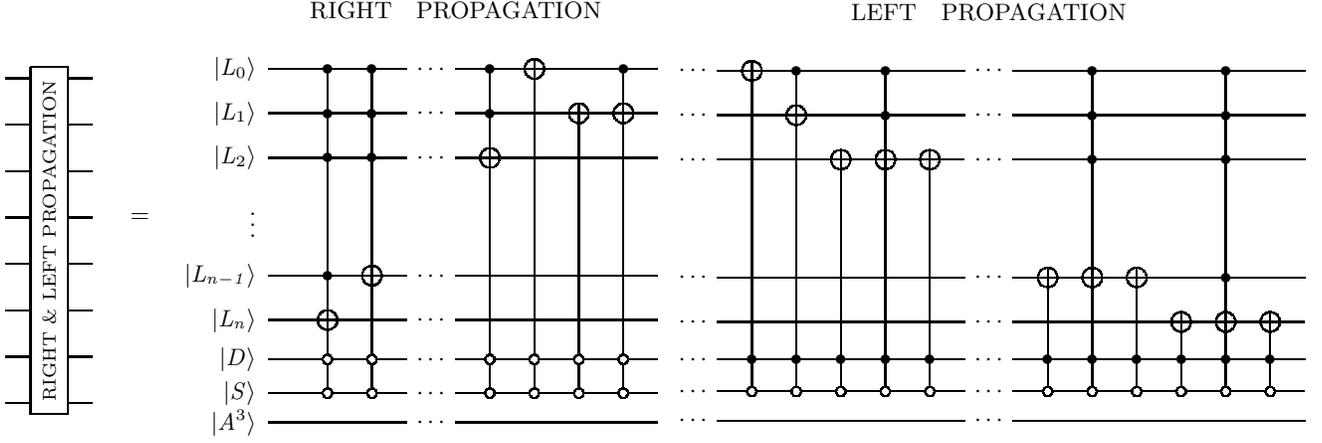
\begin{figure*}[htbp]
    \begin{tabular}{ccccc}
        \begin{minipage}[h]{0.40\linewidth}
            \centering
            \Qcircuit @C=1em @R=1em {
                & & \\
                & & \\
                & \multigate{7}{\rotatebox{90}{\scriptsize RIGHT \& LEFT PROPAGATION}} & \qw  \\
                & \ghost{1} & \qw \\
                & \ghost{1} & \qw \\
                & \ghost{1} & \qw \\
                & \ghost{1} & \qw \\
                & \ghost{1} & \qw \\ 
                & \ghost{1} & \qw \\
                & \ghost{1} & \qw 
            }
        \end{minipage} &
        \: \: = \: \: \: \: \: \: \: \: &
        \begin{minipage}[h]{0.45\linewidth}
            \centering
            \Qcircuit @C=1em @R=1em {
                & & & & & & {\rm RIGHT \quad PROPAGATION} & & & & & \\
                & & & & & & & & & & & \\
                \lstick{\ket{{\it L}_0}}     & \qw & \ctrl{1}   & \ctrl{1}   & \qw & \cdots & & \ctrl{1}   & \targ{}    & \qw        & \ctrl{1}   & \qw \\
                \lstick{\ket{{\it L}_1}}     & \qw & \ctrl{1}   & \ctrl{1}   & \qw & \cdots & & \ctrl{1}   & \qw        & \targ{}    & \targ{}    & \qw \\
                \lstick{\ket{{\it L}_2}}     & \qw & \ctrl{2}   & \ctrl{1}   & \qw & \cdots & & \targ{}    & \qw        & \qw        & \qw        & \qw \\
                & & & & & & & & & & & \\
                \lstick{\vdots} & & & & & & & & & & & \\
                & & & & & & & & & & & \\
                \lstick{\ket{{\it L_{n-1}}}} & \qw & \ctrl{-2}  & \targ{}    & \qw & \cdots & & \qw        & \qw        & \qw        & \qw        & \qw \\
                \lstick{\ket{{\it L_n}}}     & \qw & \targ{}    & \qw        & \qw & \cdots & & \qw        & \qw        & \qw        & \qw        & \qw \\
                \lstick{\ket{{\it D}}}       & \qw & \ctrlo{-3} & \ctrlo{-5} & \qw & \cdots & & \ctrlo{-6} & \ctrlo{-8} & \ctrlo{-7} & \ctrlo{-7} & \qw \\
                \lstick{\ket{{\it S}}}       & \qw & \ctrlo{-1} & \ctrlo{-1} & \qw & \cdots & & \ctrlo{-1} & \ctrlo{-1} & \ctrlo{-1} & \ctrlo{-1} & \qw \\
                \lstick{\ket{{\it A}^3}}     & \qw & \qw        & \qw        & \qw & \cdots & & \qw        & \qw        & \qw        & \qw        & \qw 
            }
        \end{minipage} &
        \begin{minipage}[h]{0.45\linewidth}
            \centering
            \Qcircuit @C=1em @R=1em {
                & & & & & & & & & {\rm LEFT \quad PROPAGATION} & & & & & & & & \\
                & & & & & & & & & & & & & & & & & \\
                & \cdots & & \targ{}    & \ctrl{1}   & \qw        & \ctrl{1}   & \qw        & \qw & \cdots & & \qw        & \ctrl{1}   & \qw        & \qw        & \ctrl{1}   & \qw        & \qw \\
                & \cdots & & \qw        & \targ{}    & \qw        & \ctrl{1}   & \qw        & \qw & \cdots & & \qw        & \ctrl{1}   & \qw        & \qw        & \ctrl{1}   & \qw        & \qw \\
                & \cdots & & \qw        & \qw        & \targ{}    & \targ{}    & \targ{}    & \qw & \cdots & & \qw        & \ctrl{1}   & \qw        & \qw        & \ctrl{1}   & \qw        & \qw \\
                & & & & & & & & & & & & & & & & & \\
                & & & & & & & & & & & & & & & & & \\
                & & & & & & & & & & & & & & & & & \\
                & \cdots & & \qw        & \qw        & \qw        & \qw        & \qw        & \qw & \cdots & & \targ{}    & \targ{}    & \targ{}    & \qw        & \ctrl{1}   & \qw        & \qw \\
                & \cdots & & \qw        & \qw        & \qw        & \qw        & \qw        & \qw & \cdots & & \qw        & \qw        & \qw        & \targ{}    & \targ{}    & \targ{}    & \qw \\
                & \cdots & & \ctrl{-8}  & \ctrl{-7}  & \ctrl{-6}  & \ctrl{-6}  & \ctrl{-6}  & \qw & \cdots & & \ctrl{-2}  & \ctrl{-5}  & \ctrl{-2}  & \ctrl{-1}  & \ctrl{-5}  & \ctrl{-1}  & \qw \\
                & \cdots & & \ctrlo{-1} & \ctrlo{-1} & \ctrlo{-1} & \ctrlo{-1} & \ctrlo{-1} & \qw & \cdots & & \ctrlo{-1} & \ctrlo{-1} & \ctrlo{-1} & \ctrlo{-1} & \ctrlo{-1} & \ctrlo{-1} & \qw \\
                & \cdots & & \qw        & \qw        & \qw        & \qw        & \qw        & \qw & \cdots & & \qw        & \qw        & \qw        & \qw        & \qw        & \qw        & \qw 
            }
        \end{minipage}
    \end{tabular}
    \caption{Quantum circuit for the right and left propagation steps.
    The symbols within the brackets are the same as in Figure \ref{circuit_absorption_step}.
    The last bracket $\ket{{\it A}^3}$ represents three ancilla qubits.}
    \label{circuit_propagation_step}
\end{figure*}

Finally, we compute the propagation step, Eq. (\ref{equation_propagation_step}), by shifting the lattice points. 
In this step, we adopt the quantum circuits for the LBM of fluid dynamics from \cite{Budinski2021}. 
The combination of CNOT gates constructs this circuit (see Figs. 5 and 6 in \cite{Budinski2021}). 
In the case of radiative transfer, a switching qubit $\ket{S}$ must be added as a control bit to shift the radiative intensity selectively. 
We illustrate this quantum circuit in Fig. \ref{circuit_propagation_step}. 
This circuit assumes periodic boundary conditions. 
This circuit changes the state $\ket{\phi_2}$ to the final state $\ket{\phi_3}$. 

After this operation, we can obtain the frequency distribution by performing measurements of all $n+5$ qubits: $\ket{L^n}$, $\ket{S}$, $\ket{D}$ and $\ket{A^3}$. 
We convert this distribution to the amplitude distribution and multiply it by a constant $2\|\phi\|$. 
Then, we obtain the radiative intensity $I_{\pm\mu}$ at $t=t+\delta_t$. 
A number of measurements is needed for this step and severely influences the efficiency of the quantum algorithm.

\section{Numerical Simulation}

In order to validate the quantum algorithm proposed in this paper, we consider a simple one-dimensional problem using a quantum circuit simulator IBM Qiskit \cite{Qiskit}. 
We set the fake backend to \texttt{Aer simulator}. 
We set $\kappa = 2.5$, $\sigma = 0.5$, and define the source term as 
\begin{eqnarray}
S_{\pm} (x) = \left\{
\renewcommand{\arraystretch}{1.5}
\begin{array}{llc}
0 & {\rm for} & 0 \leq x \leq 1/4 \\
1 & {\rm for} & 1/4 < x < 3/4 \\
0 & {\rm for} & 3/4 \leq x \leq 1
\end{array}
\right.
\end{eqnarray}
The intensity distribution reaches a steady state for $t\rightarrow\infty$.
We consider 5 lattice qubits ($i = 0 \sim 4$) and set $c = 1$, $\mu = 1$, therefore $\delta_t = \delta_x = 1/2^5$. 
We perform calculations up to $t = 2$ ($t/\delta_t=64$ time steps). 
After the propagation step, we measure the final quantum state $\ket{\phi_3}$.
In this calculation, we measure $10^6$ times for each time steps (see the Appendix). 

Fig. \ref{figure_test_problem} shows our numerical result. 
This figure also shows the result obtained using the classical method to directly solve the LBM equation (\ref{lattice_boltzmann_equation}) to compare classical numerical method to our quantum algorithm. 
The computational time $t$ and the number of lattice points $M=2^5$ are the same for both numerical calculations. 
The lines indicate the steady-state solutions derived analytically (refer to the Appendix). 
Both numerical methods reproduce the analytical solutions, which validate our quantum algorithm proposed in this paper. 
In addition, we show the time evolution of the radiative intensity distribution in Fig. \ref{figure_evolution}. 
The evolution is almost identical for both methods. 
These results indicate that our quantum algorithm can efficiently solve the LBM equation (\ref{lattice_boltzmann_equation}) as compared with the classical algorithm. 

\begin{figure}[htbp]
    \centering
    \includegraphics[scale=0.49]{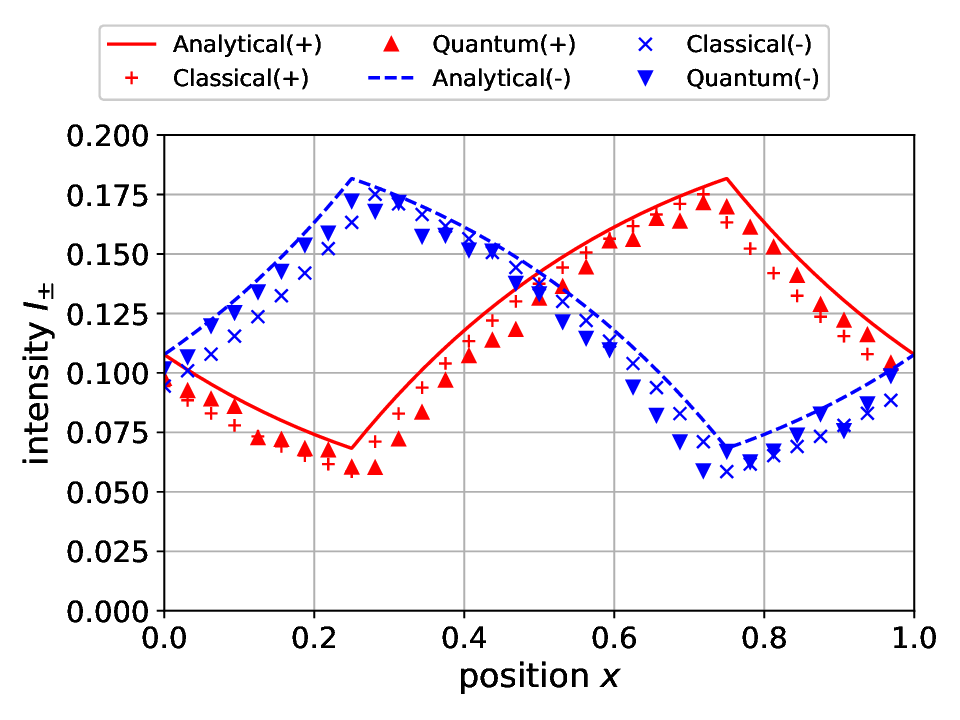}
    \caption{
    Distribution of radiative intensities $I_{\pm}$ for each method.
    Analytical represents analytical solutions, Classical denotes classical algorithm, and Quantum shows results from the Qiskit Aer simulator based on the new quantum algorithm presented in this paper.
    We show the results of numerical simulations at $t=2$.
    Refer to the main text for simulation parameters.}
    \label{figure_test_problem}
\end{figure}

\begin{figure}[htbp]
    \centering
    \includegraphics[scale=0.49]{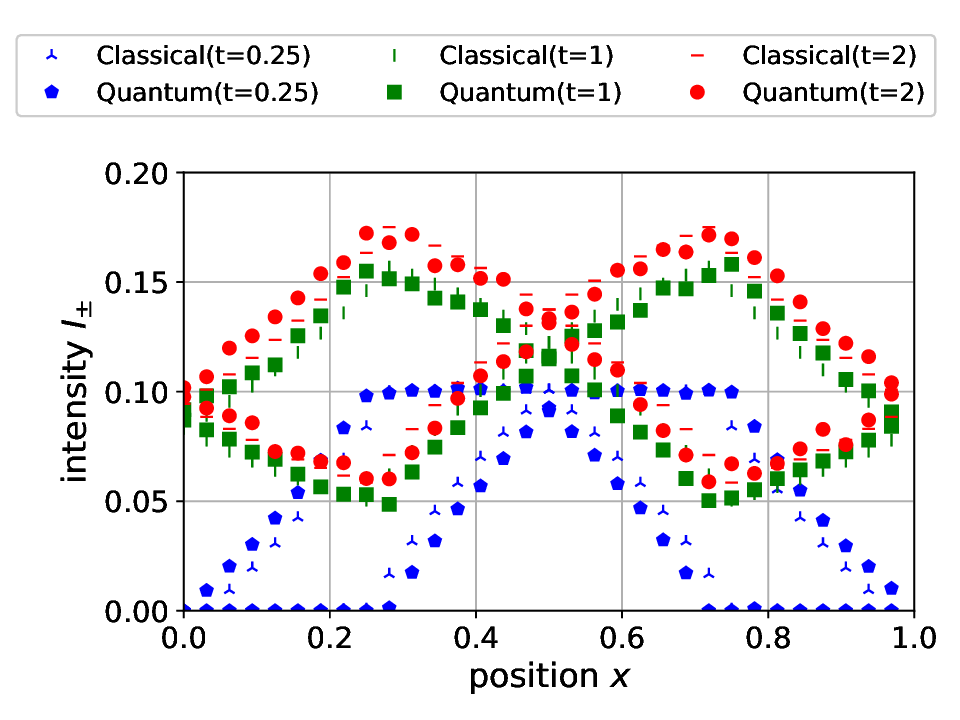}
    \caption{
    Time evolution of radiative intensities $I_{\pm}$ for each time.
    Refer to Fig. \ref{figure_test_problem} for symbols.}
    \label{figure_evolution}
\end{figure}

\section{discussion}

While this paper has addressed a one-dimensional model, we can extend it to a two-dimensional model. 
For instance, in two-dimensional space $(x,y)$, the model encompasses four directions $(\pm\mu,\pm\nu)$. 
This model is analogous to the D2Q4 model (two-dimensional flow with four discrete velocities) in the LBM of fluid dynamics. 
Equation (\ref{lattice_boltzmann_equation}) for the one-dimensional model can be extended to the two-dimensional model as
\begin{eqnarray}
&& 
I_{\pm \mu, \pm \nu} 
(x \pm \delta_x, y \pm \delta_y, t + \delta_t) 
- I_{\pm \mu, \pm \nu} (x, y, t) 
\nonumber\\
&& \quad 
= -\kappa 
I_{\pm \mu, \pm \nu} (x, y, t) 
\delta_t 
\nonumber\\
&& \qquad 
+ \frac{\sigma}{4} 
\left\{ 
I_{+\mu, +\nu} (x, y, t) 
+ I_{-\mu, +\nu} (x, y, t) 
\right. 
\nonumber\\
&& \qquad \qquad \qquad 
\left. 
+ I_{+\mu, -\nu}(x, y, t) 
+ I_{-\mu, -\nu}(x, y, t) 
\right\} 
\delta_t 
\nonumber\\
&& \qquad 
+ \frac{1}{4} 
S_{\pm \mu, \pm \nu} (x, y) 
\delta_t ,
\end{eqnarray}
The factor of 1/4 on the right-hand side represents a weight factor corresponding to the four directions. 
The number of direction qubits $\ket{D}$ is 2. 
The number of ancilla qubits $\ket{A}$ is 4. 
The switching qubit $\ket{S}$ remains the same as in the one-dimensional model. 
Moreover, to maintain the exact resolution as in a one-dimensional model, the number of lattice points $M$ becomes squared in a two-dimensional model, doubling the lattice qubits $\ket{L}$. 

In actual quantum computers, errors accumulate during the computational steps due to decoherence and noise, resulting in a solution that deviates from the exact solution. 
In order to keep the errors sufficiently small, the number of gates in the implemented quantum circuit should be minimized. 
We utilize the \texttt{transpile} function of Qiskit to estimate roughly the circuit depth used in Fig. \ref{figure_test_problem}. 
The depth depends on the Qiskit version and also the fake backends mimicking actual IBM's quantum computers. 
Because of the randomised algorithm implemented in some backends \cite{Qiskit}, the depth varies at some level. 
With the 28-qubit heavy-hex-lattice backend \texttt{FakeCambridge}, we find that the depth is around 2200 gates with a maximum of 2236 to a minimum of 2102. 
The propagation steps dominate the depth at around 1600 gates. 
In order to perform radiative transfer calculations on a real quantum computer, the reduction of gates is necessary. 
Furthermore, if we can use quantum random access memory (QRAM), it might solve the problem associated with initial encoding \cite{Giovannetti2008a, Giovannetti2008b}. 

This quantum algorithm utilizes three ancillary qubits and does not reuse them in a single time step. 
The final state $\ket{\phi_3}$ includes the unnecessary state $\ket{000^{\perp}}'_{\it a,s,d,l}$, which decreases the probability of measuring the state with $\ket{000}_{\it a}$. 
This increases the necessary number of measurements for the accuracy. 
Reusing ancillary qubits has the potential to reduce ancillary qubits and construct a more efficient quantum algorithm. 
In addition, this quantum algorithm requires measurements for each time step because of the ancillary qubits. 
If the quantum algorithm does not contain any ancillary qubits, we can perform all the time steps successively without measurements. 
Construction of the quantum algorithm without using ancillary qubits is an important future problem.

It is possible to apply radiative transfer analysis to the industrial field related to radiation phenomena. 
In heat transfer analysis, radiative heat flux becomes crucial along with thermal conduction. 
The energy equation for combined conduction-radiation heat transfer is given by
\begin{eqnarray}
&&\rho c_{\rm p} \frac{\partial T(\vec{x}, t)}{\partial t} 
= \lambda \nabla^2 T (\vec{x}, t)
- \nabla \cdot \vec{q} (\vec{x}, t),
\label{radiative_heat_transfer}
\end{eqnarray}
where $\rho$, $c_{\rm p}$, $T$, $\lambda$ and $\vec{q}$ denote density, specific heat capacity, temperature, thermal conductivity, and radiative heat flux, respectively. 
The left-hand side represents the temperature change.
The first term on the right-hand side represents thermal conduction. 
The second term indicates heat movement due to radiation, and $\vec{q}$ depends on the radiative intensity. 
Eq. (\ref{radiative_heat_transfer}) is nonlinear due to this term. 
By coupling the heat transfer equation (\ref{radiative_heat_transfer}) to the radiative transfer equation (\ref{general_radiative_transfer_equation}), one can perform radiative heat transfer analysis and determine the temperature distribution accurately. 
Conversely, since $I_{b,\nu}$ in Eq. (\ref{general_radiative_transfer_equation}) is temperature-dependent, thermal analysis is necessary for more accurate radiative transfer calculations. 
Additionally, Eq. (\ref{radiative_heat_transfer}) becomes linear without the second term on the right-hand side. 
For such a linear case, a quantum algorithm for heat transfer can be used \cite{Wei2023, Linden2022}.

\section{conclusion}

We have formulated a quantum algorithm for radiative transfer. 
In our quantum algorithm, we apply the LBM used in fluid dynamics to the radiative transfer equation. 
This equation encompasses all essential processes: absorption, scattering, and emission. 
We decompose the radiative transfer equation into three steps. 
Both the absorption-and-scattering step and the absorption-and-emission step do not act on lattice qubits (the quantum bits representing lattice points). 
Therefore in these steps, the number of gates remains unchanged even if the number of lattice points $M$ increases. 
The subsequent propagation step acts on the lattice qubits $n$. 
The depth of this operation is proportional to $n$, not to the number of lattice points ($M=2^n$). 
In classical algorithms, one must calculate for each lattice point ($M$) in the propagation step. 
Therefore, there is a potential for the quantum algorithm to accelerate calculations exponentially compared to classical algorithms. 
It is important to note that challenges remain regarding the expensive computational resources required for initial encoding (the QRAM problem) and final measurement as pointed out in previous quantum algorithms based on HHL (Harrow-Hassidim-Lloyd) algorithms \cite{Giovannetti2008a, Giovannetti2008b, Succi2023a, Harrow2009}. 

To check the validity of this quantum algorithm, we have solved a simple problem using the Qiskit Aer simulator and compared it with the exact solution. 
This test problem incorporates all essential processes in radiative transfer: absorption, scattering, and emission. 
Additionally, we compare our quantum algorithm to the classical algorithm. 
Both numerical methods well reproduce the exact solutions for radiative intensity distribution. 
Moreover, the time evolution of these distributions in both quantum and classical methods is almost identical. 
These results show the validity of this quantum algorithm based on LBM. 

Radiative transfer is essential in various fields, such as industrial engineering, nuclear fusion technology, health care, telecommunications, and astrophysics. 
Consequently, the acceleration of radiative transfer calculation using FTQCs has the potential to enhance industrial applications of FTQCs significantly.

\appendix*
\section{Analytical Solution}

In this Appendix, we derive the steady-state solutions for the test problem. 
In the absence of a source term, the system of differential equations takes the following form:
\begin{eqnarray}
     \mu \frac{\partial I_{+\mu}}{\partial x} 
    &=& -\kappa I_{+\mu} + \frac{\sigma}{2} \left( I_{+\mu} + I_{-\mu} \right), \\
    -\mu \frac{\partial I_{-\mu}}{\partial x} 
    &=& -\kappa I_{-\mu} + \frac{\sigma}{2} \left( I_{+\mu} + I_{-\mu} \right).
\end{eqnarray}
Solving these equations yields:
\begin{eqnarray}
    I_{+\mu} &=& C_+ e^{\omega x} + C_- e^{-\omega x}, \\
    I_{-\mu} &=& C_+ \left\{ \frac{2}{\sigma} (\kappa + \mu \omega) - 1 \right\} e^{\omega x} \nonumber\\
    && \quad + C_- \left\{ \frac{2}{\sigma} (\kappa - \mu \omega) - 1 \right\} e^{-\omega x}, 
\end{eqnarray}
where
\begin{eqnarray}
    \omega \equiv \sqrt{\frac{\kappa(\kappa-\sigma)}{\mu^2}},
\end{eqnarray}
and $C_+$ and $C_-$ are integration constants.
In the presence of a source term, the system of differential equations becomes:
\begin{eqnarray}
     \mu \frac{\partial I_{+\mu}}{\partial x} 
    &=& -\kappa I_{+\mu} + \frac{\sigma}{2} \left( I_{+\mu} + I_{-\mu} \right) + \frac{S_0}{2}, \\
    -\mu \frac{\partial I_{-\mu}}{\partial x} 
    &=& -\kappa I_{-\mu} + \frac{\sigma}{2} \left( I_{+\mu} + I_{-\mu} \right) + \frac{S_0}{2}.
\end{eqnarray}
Solving these equations results in
\begin{eqnarray}
    &&
    I_{+\mu} = C_1 e^{\omega x} + C_2 e^{-\omega x} + \frac{S_0}{2(\kappa - \sigma)}, \\
    &&
    I_{-\mu} = C_1 \left\{ \frac{2}{\sigma} (\kappa + \mu \omega) - 1 \right\} e^{\omega x} \nonumber\\
    && \qquad + C_2 \left\{ \frac{2}{\sigma} (\kappa - \mu \omega) - 1 \right\} e^{-\omega x} + \frac{S_0}{2(\kappa - \sigma)}, 
\end{eqnarray}
where $C_1$ and $C_2$ are integration constants.

In the test problem, we divide the system into regions with and without a source term, specifically the central region ($1/4<x<3/4$) and the non-source regions ($0<x<1/4$, $3/4<x<1$). 
Therefore, it is necessary for $I_{\pm\mu}$ to match at the boundaries of these regions ($x=1/4,3/4$). 
Additionally, due to periodic boundary conditions, $I_{\pm\mu}$ must match at both domain ends ($x=0,1$). 
These conditions allow us to determine the values of the six integration constants. 
We show the analytical solutions in Figure \ref{figure_test_problem}.

\section{Shots for Accuracy}

\begin{figure}[htbp]
    \centering
    \includegraphics[scale=0.49]{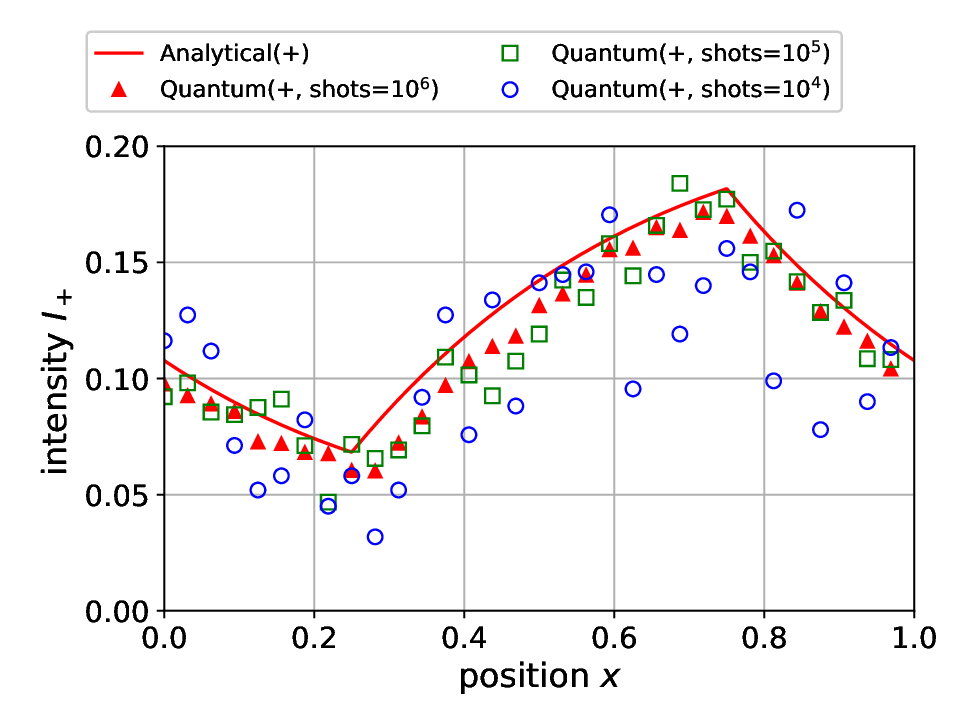}
    \caption{
    Distribution of radiative intensities $I_+$ with a variety of shots.
    The shots represent the number of measurements for each time step. 
    Refer to Fig. \ref{figure_test_problem} for symbols.
    }
    \label{figure_shots}
\end{figure}

The number of measurements (shots) should be sufficiently large to estimate the intensity distribution with a sufficient accuracy. 
In Fig. \ref{figure_shots}, we show the intensity distributions with a variety of shots. 
Fig. \ref{figure_shots} shows that the number of measurements around $10^6$ provides a sufficient accuracy. 
Also, the case of $10^5$ can give a certain degree of accuracy. 

\begin{acknowledgements}
The authors are indebted to Hiroshi Hayasaka, Takashi Imoto, Yuichiro Matsuzaki, Yuichiro Mori, and Akira Sasamoto for helpful discussions. 
This work was performed for Council for Science, Technology and Innovation (CSTI), Cross-ministerial Strategic Innovation Promotion Program (SIP), “Promoting the application of advanced quantum technology platforms to social issues”(Funding agency : QST).
\end{acknowledgements}

\nocite{*}

\bibliography{8-bibliography}

\providecommand{\noopsort}[1]{}\providecommand{\singleletter}[1]{#1}%
\begin{thebibliography}{61}%
\makeatletter
\providecommand \@ifxundefined [1]{%
 \@ifx{#1\undefined}
}%
\providecommand \@ifnum [1]{%
 \ifnum #1\expandafter \@firstoftwo
 \else \expandafter \@secondoftwo
 \fi
}%
\providecommand \@ifx [1]{%
 \ifx #1\expandafter \@firstoftwo
 \else \expandafter \@secondoftwo
 \fi
}%
\providecommand \natexlab [1]{#1}%
\providecommand \enquote  [1]{``#1''}%
\providecommand \bibnamefont  [1]{#1}%
\providecommand \bibfnamefont [1]{#1}%
\providecommand \citenamefont [1]{#1}%
\providecommand \href@noop [0]{\@secondoftwo}%
\providecommand \href [0]{\begingroup \@sanitize@url \@href}%
\providecommand \@href[1]{\@@startlink{#1}\@@href}%
\providecommand \@@href[1]{\endgroup#1\@@endlink}%
\providecommand \@sanitize@url [0]{\catcode `\\12\catcode `\$12\catcode `\&12\catcode `\#12\catcode `\^12\catcode `\_12\catcode `\%12\relax}%
\providecommand \@@startlink[1]{}%
\providecommand \@@endlink[0]{}%
\providecommand \url  [0]{\begingroup\@sanitize@url \@url }%
\providecommand \@url [1]{\endgroup\@href {#1}{\urlprefix }}%
\providecommand \urlprefix  [0]{URL }%
\providecommand \Eprint [0]{\href }%
\providecommand \doibase [0]{https://doi.org/}%
\providecommand \selectlanguage [0]{\@gobble}%
\providecommand \bibinfo  [0]{\@secondoftwo}%
\providecommand \bibfield  [0]{\@secondoftwo}%
\providecommand \translation [1]{[#1]}%
\providecommand \BibitemOpen [0]{}%
\providecommand \bibitemStop [0]{}%
\providecommand \bibitemNoStop [0]{.\EOS\space}%
\providecommand \EOS [0]{\spacefactor3000\relax}%
\providecommand \BibitemShut  [1]{\csname bibitem#1\endcsname}%
\let\auto@bib@innerbib\@empty
\bibitem [{\citenamefont {McNamara}\ and\ \citenamefont {Zanetti}(1988)}]{McNamara1988}%
  \BibitemOpen
  \bibfield  {author} {\bibinfo {author} {\bibfnamefont {G.~R.}\ \bibnamefont {McNamara}}\ and\ \bibinfo {author} {\bibfnamefont {G.}~\bibnamefont {Zanetti}},\ }\bibfield  {title} {\bibinfo {title} {Use of the {B}oltzmann equation to simulate lattice-gas automata},\ }\href@noop {} {\bibfield  {journal} {\bibinfo  {journal} {Phys. Rev. Lett.}\ }\textbf {\bibinfo {volume} {61}},\ \bibinfo {pages} {2332} (\bibinfo {year} {1988})}\BibitemShut {NoStop}%
\bibitem [{\citenamefont {Benzi}\ \emph {et~al.}(1992)\citenamefont {Benzi}, \citenamefont {Succi},\ and\ \citenamefont {Vergassola}}]{Benzi1992}%
  \BibitemOpen
  \bibfield  {author} {\bibinfo {author} {\bibfnamefont {R.}~\bibnamefont {Benzi}}, \bibinfo {author} {\bibfnamefont {S.}~\bibnamefont {Succi}},\ and\ \bibinfo {author} {\bibfnamefont {M.}~\bibnamefont {Vergassola}},\ }\bibfield  {title} {\bibinfo {title} {The lattice boltzmann equation: theory and applications},\ }\href@noop {} {\bibfield  {journal} {\bibinfo  {journal} {Physics Reports}\ }\textbf {\bibinfo {volume} {222}},\ \bibinfo {pages} {145} (\bibinfo {year} {1992})}\BibitemShut {NoStop}%
\bibitem [{\citenamefont {Higuera}\ \emph {et~al.}(1989)\citenamefont {Higuera}, \citenamefont {Succi},\ and\ \citenamefont {Benzi}}]{Higuera1989}%
  \BibitemOpen
  \bibfield  {author} {\bibinfo {author} {\bibfnamefont {F.~J.}\ \bibnamefont {Higuera}}, \bibinfo {author} {\bibfnamefont {S.}~\bibnamefont {Succi}},\ and\ \bibinfo {author} {\bibfnamefont {R.}~\bibnamefont {Benzi}},\ }\bibfield  {title} {\bibinfo {title} {Lattice gas dynamics with enhanced collisions},\ }\href@noop {} {\bibfield  {journal} {\bibinfo  {journal} {Europhysics letters}\ }\textbf {\bibinfo {volume} {9}},\ \bibinfo {pages} {345} (\bibinfo {year} {1989})}\BibitemShut {NoStop}%
\bibitem [{\citenamefont {Cali}\ \emph {et~al.}(1992)\citenamefont {Cali}, \citenamefont {Succi}, \citenamefont {Cancelliere}, \citenamefont {Benzi},\ and\ \citenamefont {Gramignani}}]{Cali1992}%
  \BibitemOpen
  \bibfield  {author} {\bibinfo {author} {\bibfnamefont {A.}~\bibnamefont {Cali}}, \bibinfo {author} {\bibfnamefont {S.}~\bibnamefont {Succi}}, \bibinfo {author} {\bibfnamefont {A.}~\bibnamefont {Cancelliere}}, \bibinfo {author} {\bibfnamefont {R.}~\bibnamefont {Benzi}},\ and\ \bibinfo {author} {\bibfnamefont {M.}~\bibnamefont {Gramignani}},\ }\bibfield  {title} {\bibinfo {title} {Diffusion and hydrodynamic dispersion with the lattice boltzmann method},\ }\href@noop {} {\bibfield  {journal} {\bibinfo  {journal} {Physical Review A}\ }\textbf {\bibinfo {volume} {45}},\ \bibinfo {pages} {5771} (\bibinfo {year} {1992})}\BibitemShut {NoStop}%
\bibitem [{\citenamefont {Qian}\ \emph {et~al.}(1992)\citenamefont {Qian}, \citenamefont {D'Humi\'{e}res},\ and\ \citenamefont {Lallemand}}]{Qian1992}%
  \BibitemOpen
  \bibfield  {author} {\bibinfo {author} {\bibfnamefont {Y.}~\bibnamefont {Qian}}, \bibinfo {author} {\bibfnamefont {D.}~\bibnamefont {D'Humi\'{e}res}},\ and\ \bibinfo {author} {\bibfnamefont {P.}~\bibnamefont {Lallemand}},\ }\bibfield  {title} {\bibinfo {title} {Lattice bgk models for {N}avier-{S}tokes equation},\ }\href@noop {} {\bibfield  {journal} {\bibinfo  {journal} {Europhysics Letter}\ }\textbf {\bibinfo {volume} {17}},\ \bibinfo {pages} {479} (\bibinfo {year} {1992})}\BibitemShut {NoStop}%
\bibitem [{\citenamefont {Alexander}\ \emph {et~al.}(1993)\citenamefont {Alexander}, \citenamefont {Chen},\ and\ \citenamefont {Sterling}}]{Alexander1993}%
  \BibitemOpen
  \bibfield  {author} {\bibinfo {author} {\bibfnamefont {F.~J.}\ \bibnamefont {Alexander}}, \bibinfo {author} {\bibfnamefont {S.}~\bibnamefont {Chen}},\ and\ \bibinfo {author} {\bibfnamefont {J.~D.}\ \bibnamefont {Sterling}},\ }\bibfield  {title} {\bibinfo {title} {Lattice {B}oltzmann thermohydrodynamics},\ }\href@noop {} {\bibfield  {journal} {\bibinfo  {journal} {Physical Review E}\ }\textbf {\bibinfo {volume} {47}},\ \bibinfo {pages} {R2249} (\bibinfo {year} {1993})}\BibitemShut {NoStop}%
\bibitem [{\citenamefont {Yan}\ \emph {et~al.}(1999)\citenamefont {Yan}, \citenamefont {Chen},\ and\ \citenamefont {Hu}}]{Yan1999}%
  \BibitemOpen
  \bibfield  {author} {\bibinfo {author} {\bibfnamefont {G.}~\bibnamefont {Yan}}, \bibinfo {author} {\bibfnamefont {Y.}~\bibnamefont {Chen}},\ and\ \bibinfo {author} {\bibfnamefont {S.}~\bibnamefont {Hu}},\ }\bibfield  {title} {\bibinfo {title} {Simple lattice {B}oltzmann model for simulating flows with shock wave},\ }\href@noop {} {\bibfield  {journal} {\bibinfo  {journal} {Physical Review E}\ }\textbf {\bibinfo {volume} {59}},\ \bibinfo {pages} {454} (\bibinfo {year} {1999})}\BibitemShut {NoStop}%
\bibitem [{\citenamefont {Kataoka}\ and\ \citenamefont {Tsutahara}(2004{\natexlab{a}})}]{Kataoka2004a}%
  \BibitemOpen
  \bibfield  {author} {\bibinfo {author} {\bibfnamefont {T.}~\bibnamefont {Kataoka}}\ and\ \bibinfo {author} {\bibfnamefont {M.}~\bibnamefont {Tsutahara}},\ }\bibfield  {title} {\bibinfo {title} {Lattice {B}oltzmann model for the compressible {N}avier-{S}tokes equations with flexible specific-heat ratio},\ }\href@noop {} {\bibfield  {journal} {\bibinfo  {journal} {Phys. Rev. E}\ }\textbf {\bibinfo {volume} {69}},\ \bibinfo {pages} {035701} (\bibinfo {year} {2004}{\natexlab{a}})}\BibitemShut {NoStop}%
\bibitem [{\citenamefont {Kataoka}\ and\ \citenamefont {Tsutahara}(2004{\natexlab{b}})}]{Kataoka2004b}%
  \BibitemOpen
  \bibfield  {author} {\bibinfo {author} {\bibfnamefont {T.}~\bibnamefont {Kataoka}}\ and\ \bibinfo {author} {\bibfnamefont {M.}~\bibnamefont {Tsutahara}},\ }\bibfield  {title} {\bibinfo {title} {Lattice {B}oltzmann method for the compressible euler equations},\ }\href@noop {} {\bibfield  {journal} {\bibinfo  {journal} {Phys. Rev. E}\ }\textbf {\bibinfo {volume} {69}},\ \bibinfo {pages} {056702} (\bibinfo {year} {2004}{\natexlab{b}})}\BibitemShut {NoStop}%
\bibitem [{\citenamefont {He}\ \emph {et~al.}(1998)\citenamefont {He}, \citenamefont {Chen},\ and\ \citenamefont {Doolen}}]{He1998}%
  \BibitemOpen
  \bibfield  {author} {\bibinfo {author} {\bibfnamefont {X.}~\bibnamefont {He}}, \bibinfo {author} {\bibfnamefont {S.}~\bibnamefont {Chen}},\ and\ \bibinfo {author} {\bibfnamefont {G.~D.}\ \bibnamefont {Doolen}},\ }\bibfield  {title} {\bibinfo {title} {A novel thermal model for the lattice {B}oltzmann method in incompressible limit},\ }\href {https://doi.org/https://doi.org/10.1006/jcph.1998.6057} {\bibfield  {journal} {\bibinfo  {journal} {Journal of Computational Physics}\ }\textbf {\bibinfo {volume} {146}},\ \bibinfo {pages} {282} (\bibinfo {year} {1998})}\BibitemShut {NoStop}%
\bibitem [{\citenamefont {Peng}\ \emph {et~al.}(2003)\citenamefont {Peng}, \citenamefont {Shu},\ and\ \citenamefont {Chew}}]{Peng2003}%
  \BibitemOpen
  \bibfield  {author} {\bibinfo {author} {\bibfnamefont {Y.}~\bibnamefont {Peng}}, \bibinfo {author} {\bibfnamefont {C.}~\bibnamefont {Shu}},\ and\ \bibinfo {author} {\bibfnamefont {Y.~T.}\ \bibnamefont {Chew}},\ }\bibfield  {title} {\bibinfo {title} {Simplified thermal lattice {B}oltzmann model for incompressible thermal flows},\ }\href {https://doi.org/10.1103/PhysRevE.68.026701} {\bibfield  {journal} {\bibinfo  {journal} {Phys. Rev. E}\ }\textbf {\bibinfo {volume} {68}},\ \bibinfo {pages} {026701} (\bibinfo {year} {2003})}\BibitemShut {NoStop}%
\bibitem [{\citenamefont {Gunstensen}\ \emph {et~al.}(1991)\citenamefont {Gunstensen}, \citenamefont {Rothman}, \citenamefont {Zaleski},\ and\ \citenamefont {Zanetti}}]{Gunstensen1991}%
  \BibitemOpen
  \bibfield  {author} {\bibinfo {author} {\bibfnamefont {A.~K.}\ \bibnamefont {Gunstensen}}, \bibinfo {author} {\bibfnamefont {D.~H.}\ \bibnamefont {Rothman}}, \bibinfo {author} {\bibfnamefont {S.}~\bibnamefont {Zaleski}},\ and\ \bibinfo {author} {\bibfnamefont {G.}~\bibnamefont {Zanetti}},\ }\bibfield  {title} {\bibinfo {title} {Lattice {B}oltzmann model of immiscible fluids},\ }\href {https://doi.org/10.1103/PhysRevA.43.4320} {\bibfield  {journal} {\bibinfo  {journal} {Phys. Rev. A}\ }\textbf {\bibinfo {volume} {43}},\ \bibinfo {pages} {4320} (\bibinfo {year} {1991})}\BibitemShut {NoStop}%
\bibitem [{\citenamefont {Shan}\ and\ \citenamefont {Chen}(1993)}]{Shan1993}%
  \BibitemOpen
  \bibfield  {author} {\bibinfo {author} {\bibfnamefont {X.}~\bibnamefont {Shan}}\ and\ \bibinfo {author} {\bibfnamefont {H.}~\bibnamefont {Chen}},\ }\bibfield  {title} {\bibinfo {title} {Lattice {B}oltzmann model for simulating flows with multiple phases and components},\ }\href {https://doi.org/10.1103/PhysRevE.47.1815} {\bibfield  {journal} {\bibinfo  {journal} {Phys. Rev. E}\ }\textbf {\bibinfo {volume} {47}},\ \bibinfo {pages} {1815} (\bibinfo {year} {1993})}\BibitemShut {NoStop}%
\bibitem [{\citenamefont {Swift}\ \emph {et~al.}(1996)\citenamefont {Swift}, \citenamefont {Orlandini}, \citenamefont {Osborn},\ and\ \citenamefont {Yeomans}}]{Swift1996}%
  \BibitemOpen
  \bibfield  {author} {\bibinfo {author} {\bibfnamefont {M.~R.}\ \bibnamefont {Swift}}, \bibinfo {author} {\bibfnamefont {E.}~\bibnamefont {Orlandini}}, \bibinfo {author} {\bibfnamefont {W.~R.}\ \bibnamefont {Osborn}},\ and\ \bibinfo {author} {\bibfnamefont {J.~M.}\ \bibnamefont {Yeomans}},\ }\bibfield  {title} {\bibinfo {title} {Lattice {B}oltzmann simulations of liquid-gas and binary fluid systems},\ }\href {https://doi.org/10.1103/PhysRevE.54.5041} {\bibfield  {journal} {\bibinfo  {journal} {Phys. Rev. E}\ }\textbf {\bibinfo {volume} {54}},\ \bibinfo {pages} {5041} (\bibinfo {year} {1996})}\BibitemShut {NoStop}%
\bibitem [{\citenamefont {Budinski}(2021)}]{Budinski2021}%
  \BibitemOpen
  \bibfield  {author} {\bibinfo {author} {\bibfnamefont {L.}~\bibnamefont {Budinski}},\ }\bibfield  {title} {\bibinfo {title} {Quantum algorithm for the advection–diffusion equation simulated with the lattice {B}oltzmann method},\ }\href@noop {} {\bibfield  {journal} {\bibinfo  {journal} {Quantum Information Processing}\ }\textbf {\bibinfo {volume} {20}},\ \bibinfo {pages} {57} (\bibinfo {year} {2021})}\BibitemShut {NoStop}%
\bibitem [{\citenamefont {Itani}\ \emph {et~al.}(2023)\citenamefont {Itani}, \citenamefont {Sreenivasan},\ and\ \citenamefont {Succi}}]{Itani2023}%
  \BibitemOpen
  \bibfield  {author} {\bibinfo {author} {\bibfnamefont {W.}~\bibnamefont {Itani}}, \bibinfo {author} {\bibfnamefont {K.~R.}\ \bibnamefont {Sreenivasan}},\ and\ \bibinfo {author} {\bibfnamefont {S.}~\bibnamefont {Succi}},\ }\href@noop {} {\bibinfo {title} {Quantum algorithm for lattice {B}oltzmann ({QALB}) simulation of incompressible fluids with a nonlinear collision term}} (\bibinfo {year} {2023}),\ \bibinfo {note} {arXiv:2304.05915}\BibitemShut {NoStop}%
\bibitem [{\citenamefont {Sanavio}\ and\ \citenamefont {Succi}(2023)}]{Sanavio2023}%
  \BibitemOpen
  \bibfield  {author} {\bibinfo {author} {\bibfnamefont {C.}~\bibnamefont {Sanavio}}\ and\ \bibinfo {author} {\bibfnamefont {S.}~\bibnamefont {Succi}},\ }\href@noop {} {\bibinfo {title} {Quantum lattice {B}oltzmann-{C}arleman algorithm}} (\bibinfo {year} {2023}),\ \bibinfo {note} {arXiv:2310.17973}\BibitemShut {NoStop}%
\bibitem [{\citenamefont {Mezzacapo}\ \emph {et~al.}(2015)\citenamefont {Mezzacapo}, \citenamefont {M.~Sanz}, \citenamefont {Egusquiza}, \citenamefont {Succi},\ and\ \citenamefont {Solano}}]{Mezzacapo2015}%
  \BibitemOpen
  \bibfield  {author} {\bibinfo {author} {\bibfnamefont {A.}~\bibnamefont {Mezzacapo}}, \bibinfo {author} {\bibfnamefont {L.~L.}\ \bibnamefont {M.~Sanz}}, \bibinfo {author} {\bibfnamefont {I.~L.}\ \bibnamefont {Egusquiza}}, \bibinfo {author} {\bibfnamefont {S.}~\bibnamefont {Succi}},\ and\ \bibinfo {author} {\bibfnamefont {E.}~\bibnamefont {Solano}},\ }\href@noop {} {\bibfield  {journal} {\bibinfo  {journal} {Scientific reports}\ }\textbf {\bibinfo {volume} {5}},\ \bibinfo {pages} {13153} (\bibinfo {year} {2015})}\BibitemShut {NoStop}%
\bibitem [{\citenamefont {Shende}\ \emph {et~al.}(2006)\citenamefont {Shende}, \citenamefont {Bullock},\ and\ \citenamefont {Markov}}]{Shende2006}%
  \BibitemOpen
  \bibfield  {author} {\bibinfo {author} {\bibfnamefont {V.}~\bibnamefont {Shende}}, \bibinfo {author} {\bibfnamefont {S.}~\bibnamefont {Bullock}},\ and\ \bibinfo {author} {\bibfnamefont {I.}~\bibnamefont {Markov}},\ }\bibfield  {title} {\bibinfo {title} {Synthesis of quantum-logic circuits},\ }\href@noop {} {\bibfield  {journal} {\bibinfo  {journal} {IEEE Trans. Comput. Aided Des. Integr. Circuits Syst.}\ }\textbf {\bibinfo {volume} {25}},\ \bibinfo {pages} {1000} (\bibinfo {year} {2006})}\BibitemShut {NoStop}%
\bibitem [{\citenamefont {Gaitan}(2020)}]{Gaitan2020}%
  \BibitemOpen
  \bibfield  {author} {\bibinfo {author} {\bibfnamefont {F.}~\bibnamefont {Gaitan}},\ }\bibfield  {title} {\bibinfo {title} {Finding flows of a {N}avier–{S}tokes fluid through quantum computing},\ }\href@noop {} {\bibfield  {journal} {\bibinfo  {journal} {npj Quantum Information}\ }\textbf {\bibinfo {volume} {6}},\ \bibinfo {pages} {61} (\bibinfo {year} {2020})}\BibitemShut {NoStop}%
\bibitem [{\citenamefont {Li}\ \emph {et~al.}(2023)\citenamefont {Li}, \citenamefont {Yin}, \citenamefont {Wiebe}, \citenamefont {Chun}, \citenamefont {Schenter}, \citenamefont {Cheung},\ and\ \citenamefont {M{\"{u}}lmenst{\"{a}}dt}}]{Li2023}%
  \BibitemOpen
  \bibfield  {author} {\bibinfo {author} {\bibfnamefont {X.}~\bibnamefont {Li}}, \bibinfo {author} {\bibfnamefont {X.}~\bibnamefont {Yin}}, \bibinfo {author} {\bibfnamefont {N.}~\bibnamefont {Wiebe}}, \bibinfo {author} {\bibfnamefont {J.}~\bibnamefont {Chun}}, \bibinfo {author} {\bibfnamefont {G.~K.}\ \bibnamefont {Schenter}}, \bibinfo {author} {\bibfnamefont {M.~S.}\ \bibnamefont {Cheung}},\ and\ \bibinfo {author} {\bibfnamefont {J.}~\bibnamefont {M{\"{u}}lmenst{\"{a}}dt}},\ }\href@noop {} {\bibinfo {title} {Potential quantum advantage for simulation of fluid dynamics}} (\bibinfo {year} {2023}),\ \bibinfo {note} {arXiv:2303.16550}\BibitemShut {NoStop}%
\bibitem [{\citenamefont {Succi}\ \emph {et~al.}(2024)\citenamefont {Succi}, \citenamefont {Itani}, \citenamefont {Sanavio}, \citenamefont {Sreenivasan},\ and\ \citenamefont {Steijl}}]{Succi2023b}%
  \BibitemOpen
  \bibfield  {author} {\bibinfo {author} {\bibfnamefont {S.}~\bibnamefont {Succi}}, \bibinfo {author} {\bibfnamefont {W.}~\bibnamefont {Itani}}, \bibinfo {author} {\bibfnamefont {C.}~\bibnamefont {Sanavio}}, \bibinfo {author} {\bibfnamefont {K.}~\bibnamefont {Sreenivasan}},\ and\ \bibinfo {author} {\bibfnamefont {R.}~\bibnamefont {Steijl}},\ }\bibfield  {title} {\bibinfo {title} {Ensemble fluid simulations on quantum computers},\ }\href@noop {} {\bibfield  {journal} {\bibinfo  {journal} {Computers \& Fluids}\ }\textbf {\bibinfo {volume} {270}},\ \bibinfo {pages} {106148} (\bibinfo {year} {2024})}\BibitemShut {NoStop}%
\bibitem [{\citenamefont {Steijl}(2019)}]{Steijl2019}%
  \BibitemOpen
  \bibfield  {author} {\bibinfo {author} {\bibfnamefont {R.}~\bibnamefont {Steijl}},\ }\bibfield  {title} {\bibinfo {title} {Quantum algorithms for fluid simulations},\ }in\ \href {https://doi.org/10.5772/intechopen.866852} {\emph {\bibinfo {booktitle} {Advances in Quantum Communication and Information}}},\ \bibinfo {editor} {edited by\ \bibinfo {editor} {\bibfnamefont {F.}~\bibnamefont {Bulnes}}, \bibinfo {editor} {\bibfnamefont {V.~N.}\ \bibnamefont {Stavrou}}, \bibinfo {editor} {\bibfnamefont {O.}~\bibnamefont {Morozov}},\ and\ \bibinfo {editor} {\bibfnamefont {A.~V.}\ \bibnamefont {Bourdine}}}\ (\bibinfo  {publisher} {IntechOpen},\ \bibinfo {address} {Rijeka},\ \bibinfo {year} {2019})\ Chap.~\bibinfo {chapter} {3}\BibitemShut {NoStop}%
\bibitem [{\citenamefont {Succi}\ \emph {et~al.}(2023)\citenamefont {Succi}, \citenamefont {Itani}, \citenamefont {Sreenivasan},\ and\ \citenamefont {Steijl}}]{Succi2023a}%
  \BibitemOpen
  \bibfield  {author} {\bibinfo {author} {\bibfnamefont {S.}~\bibnamefont {Succi}}, \bibinfo {author} {\bibfnamefont {W.}~\bibnamefont {Itani}}, \bibinfo {author} {\bibfnamefont {K.~R.}\ \bibnamefont {Sreenivasan}},\ and\ \bibinfo {author} {\bibfnamefont {R.}~\bibnamefont {Steijl}},\ }\bibfield  {title} {\bibinfo {title} {Quantum computing for ﬂuids: Where do we stand?},\ }\href {https://doi.org/doi:10.1209/0295-5075/acfdc7} {\bibfield  {journal} {\bibinfo  {journal} {Europhysics Letters}\ }\textbf {\bibinfo {volume} {144}},\ \bibinfo {pages} {10001} (\bibinfo {year} {2023})}\BibitemShut {NoStop}%
\bibitem [{\citenamefont {Harrow}\ \emph {et~al.}(2009)\citenamefont {Harrow}, \citenamefont {Hassidim},\ and\ \citenamefont {Lloyd}}]{Harrow2009}%
  \BibitemOpen
  \bibfield  {author} {\bibinfo {author} {\bibfnamefont {A.~W.}\ \bibnamefont {Harrow}}, \bibinfo {author} {\bibfnamefont {A.}~\bibnamefont {Hassidim}},\ and\ \bibinfo {author} {\bibfnamefont {S.}~\bibnamefont {Lloyd}},\ }\bibfield  {title} {\bibinfo {title} {Quantum algorithm for linear systems of equations},\ }\href@noop {} {\bibfield  {journal} {\bibinfo  {journal} {Phys. Rev. Lett.}\ }\textbf {\bibinfo {volume} {103}},\ \bibinfo {pages} {150502} (\bibinfo {year} {2009})}\BibitemShut {NoStop}%
\bibitem [{\citenamefont {Berry}(2014)}]{Berry2014}%
  \BibitemOpen
  \bibfield  {author} {\bibinfo {author} {\bibfnamefont {D.~W.}\ \bibnamefont {Berry}},\ }\bibfield  {title} {\bibinfo {title} {High-order quantum algorithm for solving linear differential equations},\ }\href@noop {} {\bibfield  {journal} {\bibinfo  {journal} {J. Phys. A Math. Theor.}\ }\textbf {\bibinfo {volume} {47}},\ \bibinfo {pages} {105301} (\bibinfo {year} {2014})}\BibitemShut {NoStop}%
\bibitem [{\citenamefont {Childs}\ \emph {et~al.}(2017)\citenamefont {Childs}, \citenamefont {Kothari}, ,\ and\ \citenamefont {Somma}}]{Childs2017}%
  \BibitemOpen
  \bibfield  {author} {\bibinfo {author} {\bibfnamefont {A.~M.}\ \bibnamefont {Childs}}, \bibinfo {author} {\bibfnamefont {R.}~\bibnamefont {Kothari}}, ,\ and\ \bibinfo {author} {\bibfnamefont {R.~D.}\ \bibnamefont {Somma}},\ }\bibfield  {title} {\bibinfo {title} {Quantum algorithm for systems of linear equations with exponentially improved dependence on precision},\ }\href@noop {} {\bibfield  {journal} {\bibinfo  {journal} {SIAM J. Comput.}\ }\textbf {\bibinfo {volume} {46}},\ \bibinfo {pages} {1920} (\bibinfo {year} {2017})}\BibitemShut {NoStop}%
\bibitem [{\citenamefont {Childs}\ \emph {et~al.}(2021)\citenamefont {Childs}, \citenamefont {Liu},\ and\ \citenamefont {Ostrander}}]{Childs2021}%
  \BibitemOpen
  \bibfield  {author} {\bibinfo {author} {\bibfnamefont {A.~M.}\ \bibnamefont {Childs}}, \bibinfo {author} {\bibfnamefont {J.-P.}\ \bibnamefont {Liu}},\ and\ \bibinfo {author} {\bibfnamefont {A.}~\bibnamefont {Ostrander}},\ }\bibfield  {title} {\bibinfo {title} {High-precision quantum algorithms for partial differential equations},\ }\href@noop {} {\bibfield  {journal} {\bibinfo  {journal} {Quantum}\ }\textbf {\bibinfo {volume} {5}},\ \bibinfo {pages} {574} (\bibinfo {year} {2021})}\BibitemShut {NoStop}%
\bibitem [{\citenamefont {Balducci}\ \emph {et~al.}(2022)\citenamefont {Balducci}, \citenamefont {Chen}, \citenamefont {Möller}, \citenamefont {Gerritsma},\ and\ \citenamefont {Breuker}}]{Balducci2022}%
  \BibitemOpen
  \bibfield  {author} {\bibinfo {author} {\bibfnamefont {G.~T.}\ \bibnamefont {Balducci}}, \bibinfo {author} {\bibfnamefont {B.}~\bibnamefont {Chen}}, \bibinfo {author} {\bibfnamefont {M.}~\bibnamefont {Möller}}, \bibinfo {author} {\bibfnamefont {M.}~\bibnamefont {Gerritsma}},\ and\ \bibinfo {author} {\bibfnamefont {R.~D.}\ \bibnamefont {Breuker}},\ }\bibfield  {title} {\bibinfo {title} {Review and perspectives in quantum computing for partial differential equations in structural mechanics},\ }\href@noop {} {\bibfield  {journal} {\bibinfo  {journal} {Front. Mech. Eng.}\ }\textbf {\bibinfo {volume} {8}} (\bibinfo {year} {2022})}\BibitemShut {NoStop}%
\bibitem [{\citenamefont {Asinari}\ \emph {et~al.}(2010)\citenamefont {Asinari}, \citenamefont {Mishra},\ and\ \citenamefont {Borchiellini}}]{Asinari2010}%
  \BibitemOpen
  \bibfield  {author} {\bibinfo {author} {\bibfnamefont {P.}~\bibnamefont {Asinari}}, \bibinfo {author} {\bibfnamefont {S.}~\bibnamefont {Mishra}},\ and\ \bibinfo {author} {\bibfnamefont {R.}~\bibnamefont {Borchiellini}},\ }\bibfield  {title} {\bibinfo {title} {A lattice {B}oltzmann formulation for the analysis of radiative heat transfer problems in a participating medium},\ }\href@noop {} {\bibfield  {journal} {\bibinfo  {journal} {Numer. Heat Transf. Part B Fundam.}\ }\textbf {\bibinfo {volume} {57}},\ \bibinfo {pages} {126} (\bibinfo {year} {2010})}\BibitemShut {NoStop}%
\bibitem [{\citenamefont {Ma}\ \emph {et~al.}(2011)\citenamefont {Ma}, \citenamefont {Dong},\ and\ \citenamefont {Tan}}]{Ma2011}%
  \BibitemOpen
  \bibfield  {author} {\bibinfo {author} {\bibfnamefont {Y.}~\bibnamefont {Ma}}, \bibinfo {author} {\bibfnamefont {S.}~\bibnamefont {Dong}},\ and\ \bibinfo {author} {\bibfnamefont {H.}~\bibnamefont {Tan}},\ }\bibfield  {title} {\bibinfo {title} {Lattice {B}oltzmann method for one-dimensional radiation transfer},\ }\href@noop {} {\bibfield  {journal} {\bibinfo  {journal} {Physical Review E}\ }\textbf {\bibinfo {volume} {84}},\ \bibinfo {pages} {016704} (\bibinfo {year} {2011})}\BibitemShut {NoStop}%
\bibitem [{\citenamefont {Bindra}\ and\ \citenamefont {Patil}(2012)}]{Bindra2012}%
  \BibitemOpen
  \bibfield  {author} {\bibinfo {author} {\bibfnamefont {H.}~\bibnamefont {Bindra}}\ and\ \bibinfo {author} {\bibfnamefont {D.~V.}\ \bibnamefont {Patil}},\ }\bibfield  {title} {\bibinfo {title} {Radiative or neutron transport modeling using a lattice {B}oltzmann equation framework},\ }\href@noop {} {\bibfield  {journal} {\bibinfo  {journal} {Physical Review E}\ }\textbf {\bibinfo {volume} {86}},\ \bibinfo {pages} {016706} (\bibinfo {year} {2012})}\BibitemShut {NoStop}%
\bibitem [{\citenamefont {Mcculloch}\ and\ \citenamefont {Bindra}(2016)}]{Mcculloch2016}%
  \BibitemOpen
  \bibfield  {author} {\bibinfo {author} {\bibfnamefont {R.}~\bibnamefont {Mcculloch}}\ and\ \bibinfo {author} {\bibfnamefont {H.}~\bibnamefont {Bindra}},\ }\bibfield  {title} {\bibinfo {title} {Coupled radiative and conjugate heat transfer in participating media using lattice {B}oltzmann methods},\ }\href@noop {} {\bibfield  {journal} {\bibinfo  {journal} {Comput. Fluids}\ }\textbf {\bibinfo {volume} {124}},\ \bibinfo {pages} {261} (\bibinfo {year} {2016})}\BibitemShut {NoStop}%
\bibitem [{\citenamefont {Gairola}\ and\ \citenamefont {Bindra}(2017)}]{Gairola2017}%
  \BibitemOpen
  \bibfield  {author} {\bibinfo {author} {\bibfnamefont {A.}~\bibnamefont {Gairola}}\ and\ \bibinfo {author} {\bibfnamefont {H.}~\bibnamefont {Bindra}},\ }\bibfield  {title} {\bibinfo {title} {Lattice {B}oltzmann method for solving non-equilibrium radiative transport problems},\ }\href@noop {} {\bibfield  {journal} {\bibinfo  {journal} {Ann. Nucl. Energy}\ }\textbf {\bibinfo {volume} {99}},\ \bibinfo {pages} {151} (\bibinfo {year} {2017})}\BibitemShut {NoStop}%
\bibitem [{\citenamefont {Zhang}\ \emph {et~al.}(2013)\citenamefont {Zhang}, \citenamefont {Yi},\ and\ \citenamefont {Tan}}]{Zhang2013}%
  \BibitemOpen
  \bibfield  {author} {\bibinfo {author} {\bibfnamefont {Y.}~\bibnamefont {Zhang}}, \bibinfo {author} {\bibfnamefont {H.}~\bibnamefont {Yi}},\ and\ \bibinfo {author} {\bibfnamefont {H.}~\bibnamefont {Tan}},\ }\bibfield  {title} {\bibinfo {title} {One-dimensional transient radiative transfer by lattice {B}oltzmann method},\ }\href@noop {} {\bibfield  {journal} {\bibinfo  {journal} {Opt. Express}\ }\textbf {\bibinfo {volume} {21}},\ \bibinfo {pages} {24532} (\bibinfo {year} {2013})}\BibitemShut {NoStop}%
\bibitem [{\citenamefont {Steijl}(2020)}]{Steijl2020}%
  \BibitemOpen
  \bibfield  {author} {\bibinfo {author} {\bibfnamefont {R.}~\bibnamefont {Steijl}},\ }\bibfield  {title} {\bibinfo {title} {Quantum algorithms for nonlinear equations in fluid mechanics},\ }in\ \href {https://doi.org/10.5772/intechopen.95023} {\emph {\bibinfo {booktitle} {Quantum Computing and Communications}}},\ \bibinfo {editor} {edited by\ \bibinfo {editor} {\bibfnamefont {Y.}~\bibnamefont {Zhao}}}\ (\bibinfo  {publisher} {IntechOpen},\ \bibinfo {address} {Rijeka},\ \bibinfo {year} {2020})\ Chap.~\bibinfo {chapter} {2}\BibitemShut {NoStop}%
\bibitem [{\citenamefont {Leyton}\ and\ \citenamefont {Osborne}(2008)}]{Leyton2008}%
  \BibitemOpen
  \bibfield  {author} {\bibinfo {author} {\bibfnamefont {S.}~\bibnamefont {Leyton}}\ and\ \bibinfo {author} {\bibfnamefont {T.}~\bibnamefont {Osborne}},\ }\href@noop {} {\bibinfo {title} {A quantum algorithm to solve nonlinear differential equations}} (\bibinfo {year} {2008}),\ \bibinfo {note} {arXiv:0812.4423}\BibitemShut {NoStop}%
\bibitem [{\citenamefont {Childs}\ and\ \citenamefont {Liu}(2020)}]{Childs2020}%
  \BibitemOpen
  \bibfield  {author} {\bibinfo {author} {\bibfnamefont {A.}~\bibnamefont {Childs}}\ and\ \bibinfo {author} {\bibfnamefont {J.-P.}\ \bibnamefont {Liu}},\ }\bibfield  {title} {\bibinfo {title} {Quantum spectral methods for differential equations},\ }\href@noop {} {\bibfield  {journal} {\bibinfo  {journal} {Comm. Math. Phys.}\ }\textbf {\bibinfo {volume} {375}},\ \bibinfo {pages} {1427} (\bibinfo {year} {2020})}\BibitemShut {NoStop}%
\bibitem [{\citenamefont {Gonz\'{a}lez-Rodr\'{i}guez}\ and\ \citenamefont {Kim}(2009)}]{Gonzalez-Rodriguez2009}%
  \BibitemOpen
  \bibfield  {author} {\bibinfo {author} {\bibfnamefont {P.}~\bibnamefont {Gonz\'{a}lez-Rodr\'{i}guez}}\ and\ \bibinfo {author} {\bibfnamefont {A.~D.}\ \bibnamefont {Kim}},\ }\bibfield  {title} {\bibinfo {title} {Comparison of light scattering models for diffuse optical tomography},\ }\href@noop {} {\bibfield  {journal} {\bibinfo  {journal} {Optics Express}\ }\textbf {\bibinfo {volume} {17}},\ \bibinfo {pages} {8756} (\bibinfo {year} {2009})}\BibitemShut {NoStop}%
\bibitem [{\citenamefont {Klose}\ and\ \citenamefont {Hielscher}(1999)}]{Klose1999}%
  \BibitemOpen
  \bibfield  {author} {\bibinfo {author} {\bibfnamefont {A.~D.}\ \bibnamefont {Klose}}\ and\ \bibinfo {author} {\bibfnamefont {A.~H.}\ \bibnamefont {Hielscher}},\ }\bibfield  {title} {\bibinfo {title} {Iterative reconstruction scheme for optical tomography based on the equation of radiative transfer},\ }\href@noop {} {\bibfield  {journal} {\bibinfo  {journal} {Med. Phys.}\ }\textbf {\bibinfo {volume} {26}},\ \bibinfo {pages} {1698} (\bibinfo {year} {1999})}\BibitemShut {NoStop}%
\bibitem [{\citenamefont {Klose}(2010)}]{Klose2010}%
  \BibitemOpen
  \bibfield  {author} {\bibinfo {author} {\bibfnamefont {A.~D.}\ \bibnamefont {Klose}},\ }\bibfield  {title} {\bibinfo {title} {The forward and inverse problem in tissue optics based on the radiative transfer equation: A brief review},\ }\href@noop {} {\bibfield  {journal} {\bibinfo  {journal} {Journal of Quantitative Spectroscopy and Radiative Transfer}\ }\textbf {\bibinfo {volume} {111}},\ \bibinfo {pages} {1852} (\bibinfo {year} {2010})}\BibitemShut {NoStop}%
\bibitem [{\citenamefont {Hoshi}\ and\ \citenamefont {Yamada}(2016)}]{Hoshi2016}%
  \BibitemOpen
  \bibfield  {author} {\bibinfo {author} {\bibfnamefont {Y.}~\bibnamefont {Hoshi}}\ and\ \bibinfo {author} {\bibfnamefont {Y.}~\bibnamefont {Yamada}},\ }\bibfield  {title} {\bibinfo {title} {Overview of diffuse optical tomography and its clinical applications},\ }\href@noop {} {\bibfield  {journal} {\bibinfo  {journal} {J. Biomed. Opt.}\ }\textbf {\bibinfo {volume} {21}},\ \bibinfo {pages} {091312} (\bibinfo {year} {2016})}\BibitemShut {NoStop}%
\bibitem [{\citenamefont {Asplund}\ \emph {et~al.}(2000)\citenamefont {Asplund}, \citenamefont {Nordlund}, \citenamefont {Trampedach}, \citenamefont {Allende},\ and\ \citenamefont {Stein}}]{Asplund2000}%
  \BibitemOpen
  \bibfield  {author} {\bibinfo {author} {\bibfnamefont {M.}~\bibnamefont {Asplund}}, \bibinfo {author} {\bibfnamefont {A.}~\bibnamefont {Nordlund}}, \bibinfo {author} {\bibfnamefont {R.}~\bibnamefont {Trampedach}}, \bibinfo {author} {\bibfnamefont {P.~C.}\ \bibnamefont {Allende}},\ and\ \bibinfo {author} {\bibfnamefont {R.~F.}\ \bibnamefont {Stein}},\ }\bibfield  {title} {\bibinfo {title} {Line formation in solar granulation. {I}. {F}e line shapes, shifts and asymmetries},\ }\href@noop {} {\bibfield  {journal} {\bibinfo  {journal} {Astronomy and Astrophysics}\ }\textbf {\bibinfo {volume} {359}},\ \bibinfo {pages} {729} (\bibinfo {year} {2000})}\BibitemShut {NoStop}%
\bibitem [{\citenamefont {Hayek}\ \emph {et~al.}(2010)\citenamefont {Hayek}, \citenamefont {Asplund}, \citenamefont {Carlsson}, \citenamefont {Trampedach}, \citenamefont {Collet}, \citenamefont {Gudiksen}, \citenamefont {Hansteen},\ and\ \citenamefont {Leenaarts}}]{Hayek2010}%
  \BibitemOpen
  \bibfield  {author} {\bibinfo {author} {\bibfnamefont {W.}~\bibnamefont {Hayek}}, \bibinfo {author} {\bibfnamefont {M.}~\bibnamefont {Asplund}}, \bibinfo {author} {\bibfnamefont {M.}~\bibnamefont {Carlsson}}, \bibinfo {author} {\bibfnamefont {R.}~\bibnamefont {Trampedach}}, \bibinfo {author} {\bibfnamefont {R.}~\bibnamefont {Collet}}, \bibinfo {author} {\bibfnamefont {B.~V.}\ \bibnamefont {Gudiksen}}, \bibinfo {author} {\bibfnamefont {V.~H.}\ \bibnamefont {Hansteen}},\ and\ \bibinfo {author} {\bibfnamefont {J.}~\bibnamefont {Leenaarts}},\ }\bibfield  {title} {\bibinfo {title} {Radiative transfer with scattering for domain-decomposed 3{D} {MHD} simulations of cool stellar atmospheres. numerical methods and application to the quiet, non-magnetic, surface of a solar-type star},\ }\href@noop {} {\bibfield  {journal} {\bibinfo  {journal} {Astronomy and Astrophysics}\ }\textbf {\bibinfo {volume} {517}},\ \bibinfo {pages} {49} (\bibinfo {year} {2010})}\BibitemShut {NoStop}%
\bibitem [{\citenamefont {Stein}\ and\ \citenamefont {Nordlund}(1998)}]{Stein1998}%
  \BibitemOpen
  \bibfield  {author} {\bibinfo {author} {\bibfnamefont {R.~F.}\ \bibnamefont {Stein}}\ and\ \bibinfo {author} {\bibfnamefont {A.}~\bibnamefont {Nordlund}},\ }\bibfield  {title} {\bibinfo {title} {Simulations of solar granulation. {I}. general properties},\ }\href@noop {} {\bibfield  {journal} {\bibinfo  {journal} {Astrophysical Journal}\ }\textbf {\bibinfo {volume} {499}},\ \bibinfo {pages} {914} (\bibinfo {year} {1998})}\BibitemShut {NoStop}%
\bibitem [{\citenamefont {Blinnikov}\ \emph {et~al.}(2000)\citenamefont {Blinnikov}, \citenamefont {Lundqvist}, \citenamefont {Bartunov}, \citenamefont {Nomoto},\ and\ \citenamefont {Iwamoto}}]{Blinnikov2000}%
  \BibitemOpen
  \bibfield  {author} {\bibinfo {author} {\bibfnamefont {S.}~\bibnamefont {Blinnikov}}, \bibinfo {author} {\bibfnamefont {P.}~\bibnamefont {Lundqvist}}, \bibinfo {author} {\bibfnamefont {O.}~\bibnamefont {Bartunov}}, \bibinfo {author} {\bibfnamefont {K.}~\bibnamefont {Nomoto}},\ and\ \bibinfo {author} {\bibfnamefont {K.}~\bibnamefont {Iwamoto}},\ }\bibfield  {title} {\bibinfo {title} {Radiation hydrodynamics of {SN} 1987{A}. {I}. global analysis of the light curve for the first 4 months},\ }\href@noop {} {\bibfield  {journal} {\bibinfo  {journal} {Astrophysical Journal}\ }\textbf {\bibinfo {volume} {532}},\ \bibinfo {pages} {1132} (\bibinfo {year} {2000})}\BibitemShut {NoStop}%
\bibitem [{\citenamefont {L.}\ \emph {et~al.}(2010)\citenamefont {L.}, \citenamefont {P.},\ and\ \citenamefont {N.}}]{Piro2010}%
  \BibitemOpen
  \bibfield  {author} {\bibinfo {author} {\bibfnamefont {P.~A.}\ \bibnamefont {L.}}, \bibinfo {author} {\bibfnamefont {C.}~\bibnamefont {P.}},\ and\ \bibinfo {author} {\bibfnamefont {W.~N.}\ \bibnamefont {N.}},\ }\bibfield  {title} {\bibinfo {title} {Shock breakout from type {I}a supernova},\ }\href@noop {} {\bibfield  {journal} {\bibinfo  {journal} {Astrophysical Journal}\ }\textbf {\bibinfo {volume} {708}},\ \bibinfo {pages} {598} (\bibinfo {year} {2010})}\BibitemShut {NoStop}%
\bibitem [{\citenamefont {Hoflich}\ and\ \citenamefont {Schaefer}(2009)}]{Hoflich2009}%
  \BibitemOpen
  \bibfield  {author} {\bibinfo {author} {\bibfnamefont {P.}~\bibnamefont {Hoflich}}\ and\ \bibinfo {author} {\bibfnamefont {B.~E.}\ \bibnamefont {Schaefer}},\ }\bibfield  {title} {\bibinfo {title} {{X}-ray and gamma-ray flashes from type {I}a supernovae?},\ }\href@noop {} {\bibfield  {journal} {\bibinfo  {journal} {Astrophysical Journal}\ }\textbf {\bibinfo {volume} {705}},\ \bibinfo {pages} {483} (\bibinfo {year} {2009})}\BibitemShut {NoStop}%
\bibitem [{\citenamefont {Noebauer}\ \emph {et~al.}(2012)\citenamefont {Noebauer}, \citenamefont {Sim}, \citenamefont {Kromer}, \citenamefont {Ropke},\ and\ \citenamefont {Hillebrandt}}]{Noebauer2012}%
  \BibitemOpen
  \bibfield  {author} {\bibinfo {author} {\bibfnamefont {U.~M.}\ \bibnamefont {Noebauer}}, \bibinfo {author} {\bibfnamefont {S.~A.}\ \bibnamefont {Sim}}, \bibinfo {author} {\bibfnamefont {M.}~\bibnamefont {Kromer}}, \bibinfo {author} {\bibfnamefont {F.~K.}\ \bibnamefont {Ropke}},\ and\ \bibinfo {author} {\bibfnamefont {W.}~\bibnamefont {Hillebrandt}},\ }\bibfield  {title} {\bibinfo {title} {{M}onte {C}arlo radiation hydrodynamics: methods, tests and application to type {I}a supernova ejecta},\ }\href@noop {} {\bibfield  {journal} {\bibinfo  {journal} {Monthly Notices of the Royal Astronomical Society}\ }\textbf {\bibinfo {volume} {425}},\ \bibinfo {pages} {1430} (\bibinfo {year} {2012})}\BibitemShut {NoStop}%
\bibitem [{\citenamefont {Kasen}(2010)}]{Kasen2010}%
  \BibitemOpen
  \bibfield  {author} {\bibinfo {author} {\bibfnamefont {D.}~\bibnamefont {Kasen}},\ }\bibfield  {title} {\bibinfo {title} {Seeing the collision of a supernova with its companion star},\ }\href@noop {} {\bibfield  {journal} {\bibinfo  {journal} {Astrophysical Journal}\ }\textbf {\bibinfo {volume} {708}},\ \bibinfo {pages} {1025} (\bibinfo {year} {2010})}\BibitemShut {NoStop}%
\bibitem [{\citenamefont {Fryer}\ \emph {et~al.}(2010)\citenamefont {Fryer}, \citenamefont {Ruiter}, \citenamefont {Belczynski}, \citenamefont {Brown}, \citenamefont {Bufano}, \citenamefont {Diehl}, \citenamefont {Fontes}, \citenamefont {Frey}, \citenamefont {Holland}, \citenamefont {Hungerford}, \citenamefont {Immler}, \citenamefont {Mazzali}, \citenamefont {Meakin}, \citenamefont {Milne}, \citenamefont {Raskin},\ and\ \citenamefont {Timmes}}]{Fryer2010}%
  \BibitemOpen
  \bibfield  {author} {\bibinfo {author} {\bibfnamefont {C.~L.}\ \bibnamefont {Fryer}}, \bibinfo {author} {\bibfnamefont {A.~J.}\ \bibnamefont {Ruiter}}, \bibinfo {author} {\bibfnamefont {K.}~\bibnamefont {Belczynski}}, \bibinfo {author} {\bibfnamefont {P.~J.}\ \bibnamefont {Brown}}, \bibinfo {author} {\bibfnamefont {F.}~\bibnamefont {Bufano}}, \bibinfo {author} {\bibfnamefont {S.}~\bibnamefont {Diehl}}, \bibinfo {author} {\bibfnamefont {C.~J.}\ \bibnamefont {Fontes}}, \bibinfo {author} {\bibfnamefont {L.~H.}\ \bibnamefont {Frey}}, \bibinfo {author} {\bibfnamefont {S.~T.}\ \bibnamefont {Holland}}, \bibinfo {author} {\bibfnamefont {A.~L.}\ \bibnamefont {Hungerford}}, \bibinfo {author} {\bibfnamefont {S.}~\bibnamefont {Immler}}, \bibinfo {author} {\bibfnamefont {P.}~\bibnamefont {Mazzali}}, \bibinfo {author} {\bibfnamefont {C.}~\bibnamefont {Meakin}}, \bibinfo {author} {\bibfnamefont {P.~A.}\ \bibnamefont {Milne}}, \bibinfo {author} {\bibfnamefont {C.}~\bibnamefont {Raskin}},\ and\ \bibinfo {author}
  {\bibfnamefont {F.~X.}\ \bibnamefont {Timmes}},\ }\bibfield  {title} {\bibinfo {title} {Spectra of type {I}a supernovae from double degenerate mergers},\ }\href@noop {} {\bibfield  {journal} {\bibinfo  {journal} {Astrophysical Journal}\ }\textbf {\bibinfo {volume} {725}},\ \bibinfo {pages} {296} (\bibinfo {year} {2010})}\BibitemShut {NoStop}%
\bibitem [{\citenamefont {Yahia}\ \emph {et~al.}(2021)\citenamefont {Yahia}, \citenamefont {Meraihi}, \citenamefont {Ramdane-Cherif}, \citenamefont {G.}, \citenamefont {Acheli},\ and\ \citenamefont {Guan}}]{Yahia2021}%
  \BibitemOpen
  \bibfield  {author} {\bibinfo {author} {\bibfnamefont {S.}~\bibnamefont {Yahia}}, \bibinfo {author} {\bibfnamefont {Y.}~\bibnamefont {Meraihi}}, \bibinfo {author} {\bibfnamefont {A.}~\bibnamefont {Ramdane-Cherif}}, \bibinfo {author} {\bibfnamefont {A.~B.}\ \bibnamefont {G.}}, \bibinfo {author} {\bibfnamefont {D.}~\bibnamefont {Acheli}},\ and\ \bibinfo {author} {\bibfnamefont {H.}~\bibnamefont {Guan}},\ }\bibfield  {title} {\bibinfo {title} {A survey of channel modeling techniques for visible light communications},\ }\href {https://doi.org/https://doi.org/10.1016/j.jnca.2021.103206} {\bibfield  {journal} {\bibinfo  {journal} {Journal of Network and Computer Applications}\ }\textbf {\bibinfo {volume} {194}},\ \bibinfo {pages} {103206} (\bibinfo {year} {2021})}\BibitemShut {NoStop}%
\bibitem [{\citenamefont {Gonome}\ \emph {et~al.}(2014)\citenamefont {Gonome}, \citenamefont {Baneshi}, \citenamefont {Okajima}, \citenamefont {Komiya}, \citenamefont {Yamada},\ and\ \citenamefont {Maruyama}}]{Gonome2014}%
  \BibitemOpen
  \bibfield  {author} {\bibinfo {author} {\bibfnamefont {H.}~\bibnamefont {Gonome}}, \bibinfo {author} {\bibfnamefont {M.}~\bibnamefont {Baneshi}}, \bibinfo {author} {\bibfnamefont {J.}~\bibnamefont {Okajima}}, \bibinfo {author} {\bibfnamefont {A.}~\bibnamefont {Komiya}}, \bibinfo {author} {\bibfnamefont {N.}~\bibnamefont {Yamada}},\ and\ \bibinfo {author} {\bibfnamefont {S.}~\bibnamefont {Maruyama}},\ }\bibfield  {title} {\bibinfo {title} {Control of thermal barrier performance by optimized nanoparticle size and experimental evaluation using a solar simulator},\ }\href {https://doi.org/https://doi.org/10.1016/j.jqsrt.2014.07.025} {\bibfield  {journal} {\bibinfo  {journal} {Journal of Quantitative Spectroscopy and Radiative Transfer}\ }\textbf {\bibinfo {volume} {149}},\ \bibinfo {pages} {81} (\bibinfo {year} {2014})}\BibitemShut {NoStop}%
\bibitem [{\citenamefont {Su}\ \emph {et~al.}(2017)\citenamefont {Su}, \citenamefont {Min}, \citenamefont {Cao}, \citenamefont {Sun}, \citenamefont {Hayden}, \citenamefont {O’Sullivan},\ and\ \citenamefont {Dong}}]{Su2017}%
  \BibitemOpen
  \bibfield  {author} {\bibinfo {author} {\bibfnamefont {M.~G.}\ \bibnamefont {Su}}, \bibinfo {author} {\bibfnamefont {Q.}~\bibnamefont {Min}}, \bibinfo {author} {\bibfnamefont {S.~Q.}\ \bibnamefont {Cao}}, \bibinfo {author} {\bibfnamefont {D.~X.}\ \bibnamefont {Sun}}, \bibinfo {author} {\bibfnamefont {P.}~\bibnamefont {Hayden}}, \bibinfo {author} {\bibfnamefont {G.}~\bibnamefont {O’Sullivan}},\ and\ \bibinfo {author} {\bibfnamefont {C.~Z.}\ \bibnamefont {Dong}},\ }\bibfield  {title} {\bibinfo {title} {Evolution analysis of {EUV} radiation from laser-produced tin plasmas based on a radiation hydrodynamics model},\ }\href {https://doi.org/10.1038/srep45212} {\bibfield  {journal} {\bibinfo  {journal} {Scientific Reports}\ }\textbf {\bibinfo {volume} {7}},\ \bibinfo {pages} {45212} (\bibinfo {year} {2017})}\BibitemShut {NoStop}%
\bibitem [{\citenamefont {Chandrasekhar}(2011)}]{Chandrasekhar2011}%
  \BibitemOpen
  \bibfield  {author} {\bibinfo {author} {\bibfnamefont {S.}~\bibnamefont {Chandrasekhar}},\ }\href@noop {} {\emph {\bibinfo {title} {Radiative Transfer}}}\ (\bibinfo  {publisher} {Dover Publications},\ \bibinfo {year} {2011})\BibitemShut {NoStop}%
\bibitem [{\citenamefont {Succi}(2001)}]{Succi2001}%
  \BibitemOpen
  \bibfield  {author} {\bibinfo {author} {\bibfnamefont {S.}~\bibnamefont {Succi}},\ }\href@noop {} {\emph {\bibinfo {title} {The Lattice {B}oltzmann Equation for Fluid Dynamics and Beyond}}}\ (\bibinfo  {publisher} {Clarendon Press},\ \bibinfo {year} {2001})\BibitemShut {NoStop}%
\bibitem [{\citenamefont {{Qiskit contributors}}(2023)}]{Qiskit}%
  \BibitemOpen
  \bibfield  {author} {\bibinfo {author} {\bibnamefont {{Qiskit contributors}}},\ }\href@noop {} {\bibinfo {title} {{Q}iskit: An open-source framework for quantum computing}} (\bibinfo {year} {2023}),\ \bibinfo {note} {10.5281/zenodo.2573505}\BibitemShut {NoStop}%
\bibitem [{\citenamefont {Giovannetti}\ \emph {et~al.}(2008{\natexlab{a}})\citenamefont {Giovannetti}, \citenamefont {Lloyd},\ and\ \citenamefont {Maccone}}]{Giovannetti2008a}%
  \BibitemOpen
  \bibfield  {author} {\bibinfo {author} {\bibfnamefont {V.}~\bibnamefont {Giovannetti}}, \bibinfo {author} {\bibfnamefont {S.}~\bibnamefont {Lloyd}},\ and\ \bibinfo {author} {\bibfnamefont {L.}~\bibnamefont {Maccone}},\ }\bibfield  {title} {\bibinfo {title} {Architectures for a quantum random access memory},\ }\href@noop {} {\bibfield  {journal} {\bibinfo  {journal} {Phys. Rev. A}\ }\textbf {\bibinfo {volume} {78}},\ \bibinfo {pages} {052310} (\bibinfo {year} {2008}{\natexlab{a}})}\BibitemShut {NoStop}%
\bibitem [{\citenamefont {Giovannetti}\ \emph {et~al.}(2008{\natexlab{b}})\citenamefont {Giovannetti}, \citenamefont {Lloyd},\ and\ \citenamefont {Maccone}}]{Giovannetti2008b}%
  \BibitemOpen
  \bibfield  {author} {\bibinfo {author} {\bibfnamefont {V.}~\bibnamefont {Giovannetti}}, \bibinfo {author} {\bibfnamefont {S.}~\bibnamefont {Lloyd}},\ and\ \bibinfo {author} {\bibfnamefont {L.}~\bibnamefont {Maccone}},\ }\bibfield  {title} {\bibinfo {title} {Quantum random access memory},\ }\href@noop {} {\bibfield  {journal} {\bibinfo  {journal} {Phys. Rev. Lett.}\ }\textbf {\bibinfo {volume} {100}},\ \bibinfo {pages} {160501} (\bibinfo {year} {2008}{\natexlab{b}})}\BibitemShut {NoStop}%
\bibitem [{\citenamefont {Wei}\ \emph {et~al.}(2023)\citenamefont {Wei}, \citenamefont {Wei}, \citenamefont {Lu}, \citenamefont {Shao}, \citenamefont {Gao}, \citenamefont {Zhou}, \citenamefont {Li}, \citenamefont {Xin},\ and\ \citenamefont {Long}}]{Wei2023}%
  \BibitemOpen
  \bibfield  {author} {\bibinfo {author} {\bibfnamefont {S.~J.}\ \bibnamefont {Wei}}, \bibinfo {author} {\bibfnamefont {C.}~\bibnamefont {Wei}}, \bibinfo {author} {\bibfnamefont {P.}~\bibnamefont {Lu}}, \bibinfo {author} {\bibfnamefont {C.}~\bibnamefont {Shao}}, \bibinfo {author} {\bibfnamefont {P.}~\bibnamefont {Gao}}, \bibinfo {author} {\bibfnamefont {Z.}~\bibnamefont {Zhou}}, \bibinfo {author} {\bibfnamefont {K.}~\bibnamefont {Li}}, \bibinfo {author} {\bibfnamefont {T.}~\bibnamefont {Xin}},\ and\ \bibinfo {author} {\bibfnamefont {G.~L.}\ \bibnamefont {Long}},\ }\bibfield  {title} {\bibinfo {title} {A quantum algorithm for heat conduction with symmetrization},\ }\href@noop {} {\bibfield  {journal} {\bibinfo  {journal} {Science Bulletin}\ }\textbf {\bibinfo {volume} {68}},\ \bibinfo {pages} {494} (\bibinfo {year} {2023})}\BibitemShut {NoStop}%
\bibitem [{\citenamefont {Linden}\ \emph {et~al.}(2022)\citenamefont {Linden}, \citenamefont {Montanaro},\ and\ \citenamefont {Shao}}]{Linden2022}%
  \BibitemOpen
  \bibfield  {author} {\bibinfo {author} {\bibfnamefont {N.}~\bibnamefont {Linden}}, \bibinfo {author} {\bibfnamefont {A.}~\bibnamefont {Montanaro}},\ and\ \bibinfo {author} {\bibfnamefont {C.}~\bibnamefont {Shao}},\ }\bibfield  {title} {\bibinfo {title} {Quantum vs. classical algorithms for solving the heat equation},\ }\href@noop {} {\bibfield  {journal} {\bibinfo  {journal} {Commun. Math. Phys.}\ }\textbf {\bibinfo {volume} {395}},\ \bibinfo {pages} {601} (\bibinfo {year} {2022})}\BibitemShut {NoStop}%
\end{thebibliography}%

\end{document}